\documentclass[float,aps,prc,superscriptaddress,showpacs,twocolumn]{revtex4-1}
\usepackage{amsmath}
\usepackage{bm}
\usepackage{color}
\usepackage{graphicx,epsfig}
\usepackage{longtable}
\usepackage{amssymb}

\usepackage{soul}

\newcommand{\nuc}[2] {$^{#1}$#2}

\def\beq{\begin{equation}}
\def\eeq{\end{equation}}
\def\beqy{\begin{eqnarray}}
\def\eeqy{\end{eqnarray}}
\newcommand{\bra}{{\langle}}
\newcommand{\ket}{{\rangle}}
\newcommand{\ev}[1]{{\bra {#1} \ket}}

\newcommand{\bs}[1]{\ensuremath{\boldsymbol{#1}}}

\newcommand{\rmsPP}{{1.96734(1)}}
\newcommand{\rmsFF}{{2.1219(1)}}
\newcommand{\rfourthPP}{{55.370(1)}}
\newcommand{\rfourthFF}{{64.809(1)}}
\newcommand{\thirdZemPP}{{31.7812(3)}}
\newcommand{\thirdZemFF}{{38.2902(3)}}
\newcommand{\rzPP}{{2.3811(2)}}
\newcommand{\rzFF}{{2.5973(2)}}
\newcommand{\rmagPP}{{1.9405(1)}}
\newcommand{\rmagFF}{{2.0664(1)}}

\begin{document}

\title{Zemach moments and radii of $^{2,3}$H and $^{3,4}$He}

\author{N.~Nevo Dinur}
\affiliation{TRIUMF, 4004 Wesbrook Mall, Vancouver, BC, V6T 2A3, Canada}
\author{O.J.~Hernandez}
\affiliation{TRIUMF, 4004 Wesbrook Mall, Vancouver, BC, V6T 2A3, Canada}
\affiliation{Department of Physics and Astronomy, University of British Columbia, Vancouver, BC, V6T 1Z4, Canada}
\affiliation{Institut f\"{u}r Kernphysik and PRISMA Cluster of Excellence, Johannes-Gutenberg Universit\"{a}t, Mainz, DE-55128, Germany}
\author{S.~Bacca}\affiliation{Institut f\"{u}r Kernphysik and PRISMA Cluster of Excellence, Johannes-Gutenberg Universit\"{a}t, Mainz, DE-55128, Germany}
\affiliation{TRIUMF, 4004 Wesbrook Mall, Vancouver, BC, V6T 2A3, Canada}
\affiliation{Department of Physics and Astronomy, University of Manitoba, Winnipeg, MB, R3T 2N2, Canada}
\author{N.~Barnea}\affiliation{Racah Institute of Physics, The Hebrew University, Jerusalem 91904, Israel}
\author{C.~Ji}
\affiliation{Key Laboratory of Quark and Lepton Physics (MOE) and Institute of Particle Physics, Central China Normal University, Wuhan 430079, China}
\author{S.~Pastore}
\affiliation{ Physics Department, Washington University, St Louis, MO 63130, USA}
\author{M.~Piarulli}
\affiliation{ Physics Department, Washington University, St Louis, MO 63130, USA}
\author{R.B.~Wiringa}
\affiliation{Physics Division, Argonne National Laboratory, Argonne, IL 60439, USA}

\date{\today}

\begin{abstract}
We present  benchmark  calculations of Zemach moments and radii of $^{2,3}$H and $^{3,4}$He 
using various few-body methods. Zemach moments are required to interpret muonic atom data 
measured by the CREMA collaboration at the Paul Scherrer Institute. Conversely,  radii extracted
from spectroscopic measurements can be compared with ab initio computations, posing stringent
constraints on the nuclear model. For a given few-body method, 
different numerical  procedures can be applied to compute these quantities. 
A detailed analysis of the numerical uncertainties entering the  
total theoretical error is presented. Uncertainties from the few-body method and the 
calculational procedure are found to be smaller than the dependencies on the 
dynamical modeling and the single nucleon inputs, which are found to be $\lesssim 2\%$. 
When relativistic corrections and two-body currents are accounted for, the 
calculated moments and radii are in very good agreement with the available experimental data. 
\end{abstract}


\maketitle

%%%%%%%%%%%%%%%%%%%%%%%%%%%%%%%%%%%%%%%%%%%%%%%%%
\section {Introduction}
\label{sec:intro}

Recent spectroscopic measurements on muonic atoms have enabled an extraction 
of the charge radii of the proton~\cite{Pohl:2010zza,Antognini13} and 
deuteron~\cite{Pohl_2016_Science} with unprecedented precision, exposing 
inconsistencies with measurements performed on electronic systems, 
see, e.g., Refs.~\cite{Metrologia2017,Bernauer_Pohl,Pohl_Review,Beyer79,Paris}. 
The emergence of the so-called ``proton-radius'' and ``deuteron-radius'' 
puzzles has attracted the attention of both the experimental and theoretical 
physics communities. Regardless of the nature of these puzzles,
it became clear that the precise determination of any nuclear charge radius 
from spectroscopic measurements on its muonic atom/ion heavily relies 
on an accurate knowledge of nuclear structure corrections to the 
muonic spectrum~\cite{Pohl_2016_LEAP,Ji:2018}. 
The CREMA collaboration has began investigating other light systems, 
such as $^{3,4}$He~\cite{Pohl_2016_LEAP,Franke_2017,Diepold_2018}, therefore,
detailed studies on light nuclei are called for, and  demand a careful investigation 
of all sources of uncertainty.

In a hydrogen-like muonic atom or ion, 
the energy difference $2S$--$2P$, also called the Lamb shift (LS), is a sensitive 
probe of the charge distribution of the nucleus (see, e.g., 
Refs.~\cite{Eides_PR2001,Borie2012_AoP_arXiv} for reviews and 
Ref.~\cite{Korzinin_2018} and references therein for the most recent calculations). 
In a $Z\alpha$ expansion up to 5$^{\rm th}$ order, with $\alpha$ being the fine-structure
constant and $Z$ the proton number, this energy shift is related to the rms electric charge radius of the nucleus 
$R_{E}\equiv\sqrt{\bra R_{E}^2\ket}$ 
by
\begin{equation} 
\label{eq:E2s2p}
\Delta E_{\rm LS}  =
\delta_{\rm QED}+
% \frac{m_r^3 (Z \alpha)^4}{12} 
\mathcal{A}_{\rm QED} \cdot
R_{E}^2 + \delta_{\rm TPE}\,, 
\end{equation}
where 
$\delta_{\rm QED}$ 
and $\mathcal{A}_{\rm QED}$ 
are independent of nuclear structure  and 
are known to a very high accuracy from quantum electro-dynamics (QED).
The precision of the radius extracted from 
these measurements is  driven by the uncertainty in the  $\delta_{\rm TPE}$ term~\cite{Ji:2018}. 
The latter describes the two-photon 
exchange (TPE) process where two virtual photons transfer energy and momentum to and from the nucleus.
We note in passing that an analogous expression to Eq.~\eqref{eq:E2s2p} 
allows the extraction of the Zemach radius $\bra R_Z \ket$ (defined in Section~\ref{sec:Numerical}) 
from the measured hyperfine splitting (HFS) of
a muonic $nS$ states~\cite{Pohl_2016_LEAP}. Also in this case, accurate 
nuclear structure calculations of the TPE contribution play a crucial 
role~\cite{Marcin}.

The $\delta_{\rm TPE}$ term can be separated into elastic and inelastic 
contributions. In the second case, the nucleus is excited to intermediate states.
The elastic contribution $\delta_{\rm Zem}$ is  related to the third Zemach 
moment of the electric form factor $\ev{R^3_E}_{(2)}$, 
while the inelastic term $\delta_{\rm pol}$ is related to the nuclear polarizability, 
so that $\delta_{\rm TPE}= \delta_{\rm Zem}+\delta_{\rm pol}$.
Notably, ab initio calculations reported in Ref.~\cite{Ji:2018} currently provide the most 
precise determinations of $\delta_{\rm pol}$ and 
$\delta_{\rm Zem}$ values and
include nucleons' finite sizes but neglect 
the contributions from two-body currents.
One of the goals of this paper is to study the effect of two-body currents, 
nuclear models, and different treatments of single-nucleon finite-sizes.

The puzzles exposed by muonic laser spectroscopy have contributed 
to the evolution of nuclear theory into a new era of precision,  where the 
various sources of theoretical uncertainty need to be addressed adequately. 
It is worth noting that while there has been considerable activity recently 
devoted to the theoretical evaluation of $\delta_{\rm TPE}$ in light muonic 
atoms~\cite{Pachucki2011,Friar:2013rha,Ji13,Carlson:2013xea,Hernandez14,Pachucki2015,NND16,Carlson:2016cii}, 
the variety of few-body methods used in ab initio nuclear physics have yet to confront the computation 
of nuclear Zemach moments and similar observables. A famous benchmark of different few-body methods for 
computing the binding energy and radius of $^4$He dates back almost two decades~\cite{Benchmark_2001} and 
thus did not utilize state-of-the-art nuclear forces. More recently, other four-body and even five-body 
benchmarks were performed, e.g., in Refs.~\cite{Benchmark_Hadronic,Lazauskas:2017paz}, which focused on 
hadronic scattering rather than on electromagnetic observables.
Filling this gap is among the goals of this work. To this end, we benchmark
different ab initio methods on electromagnetic radii, Zemach moments, and 
other ground-state observables for light nuclei in the mass range of $2\le A\le 4$, 
which are of interest to the ongoing experimental efforts mentioned above.

We focus on ground-state observables that can be readily calculated by 
the few-body methods adopted here. We  neglect
$\delta_{\rm pol}$---which requires an additional computational development, 
as described in Refs.~\cite{Ji:2018,NND_2014_PRC,Baker_2018}.

In particular,  we solve the  $A=2$ problem using either the Numerov algorithm~\cite{Numerov} 
or the harmonic oscillator expansion used in Refs.~\cite{Hernandez14,Hernandez2018,Ji:2018}.
For $A=3$ and 4, we use  Variational Monte Carlo (VMC)~\cite{Wiringa91} and Green's function Monte Carlo 
(GFMC)~\cite{pudliner1997} methods, along with two different implementations of
the  hyperspherical harmonics (HH) expansions, namely its momentum-space formulation 
(HH-p)~\cite{Piarulli12} and the effective interaction scheme in coordinate-space (EIHH)~\cite{Barnea2001}. 
These are all well-established methods and we do not provide further details on them here, 
but rather refer the interested reader to the following 
articles~\cite{Leidemann12,review2014,Marcucci:2015rca,Carlson:1997qn,Kievsky:2008es,Viviani:2005gu}.

The paper is structured as follows. In Section~\ref{sec:Numerical}, we define the 
various electromagnetic observables under study and present the numerical procedures 
implemented for their computations.
In Section~\ref{sec:Results}, we perform a benchmark in the impulse approximation (IA) using 
wave functions from different few-body methods for $A=2,3$ and $4$ systems,  and compare the results 
to experimental data. The agreement with data is reached by including  relativistic corrections 
and two-body currents, whose contributions are studied only for the $A=3$ nuclei. 
Finally, we probe the sensitivity of our results to variations in both nuclear and nucleonic inputs,
and in Section~\ref{sec:Conclusion} we draw our conclusions.

%%%%%%%%%%%%%%%%%%%%%%%%%%%%%%%%%%%%%%%%%%%%%%%%%
\section {Numerical procedures}
\label{sec:Numerical}

For a given few-body method that can provide nuclear wave functions, 
different procedures can be used to calculate both Zemach and regular electromagnetic moments.
We present a momentum-space, a coordinate-space, and a mixed (momentum \& coordinate-spaces)
formulation.  The latter exploits the respective advantages of the previous two methods.

\subsection{Definitions and momentum-space formulation}
The electric ($E$) and magnetic ($M$) form factors are 
defined in momentum-space~\cite{Piarulli12,DanielAR}
as expectation values of the  ground state wave function 
of the A-body nucleus. In particular, the deuteron electric 
and magnetic form factors are defined, respectively, as~\cite{Piarulli12}
\begin{eqnarray}
F_E(q^2)&=&\frac{1}{3} \sum_{M=\pm 1,0} \langle d;M\mid \rho(q\,\hat{\bf z})
\mid d;M\rangle \ ,   \\
F_M(q^2)&=&   \frac{\sqrt{2}\,m_d}{q} \,{\rm Im}\left[\,  \langle d;1\mid j_y(q\,\hat{\bf z})
\mid  d;0\rangle \, \right]\ , 
\end{eqnarray}
where $\mid d;M\rangle$ is the deuteron state with spin projection
$J_z=M$, $\rho$ and $j_y$ denote, respectively, the charge
operator and $y$-component of the current operator; the momentum transfer ${\bf q}$ is
taken along the $z$-axis (the spin quantization axis), and $m_d$ is the deuteron mass. 
Form factors are normalized at $q^2=0$ to $1$ and $(m_d/m_N) \mu_d$, 
respectively, where $\mu_d$ is the deuteron magnetic moment (in units of nucleon Bohr magneton $\mu_N$).

The charge and magnetic form factors of the trinucleons are 
derived from~\cite{Piarulli12} 
\begin{eqnarray}
\label{ff1}
F_E(q^2) &=& \frac{1}{Z} \, \langle +\! \mid \rho(q\, \hat{\bf z}) \mid\! + \rangle \ , \\
F_M(q^2) &=& -\frac{2\, m_N}{q} \, 
{\rm Im} \left [ \,\langle -\!\mid j_y(q\, \hat{\bf z}) \mid\!  + \rangle \, \right] \ ,
\label{ff2}
\end{eqnarray}
normalized to $1$ and $\mu$,
where $\mu$ is the magnetic moment  of the three-body system (in units of $\mu_N$), and $\mid \pm\rangle$
represent either the $^3$He state or $^3$H state in spin projections $J_z=\pm 1/2$.

The charge and current operators are  expanded in many-body terms as
\begin{eqnarray}
  \label{operators}
\rho(q)&=& \sum_i^A {\rho}_i (q) + \sum_{i<j}^A {\rho}_{ij}(q) + \dots\ , \\
\nonumber
{\bf j}(q) &=&  \sum_i^A {\bf j}_i(q) + \sum_{i<j}^A {\bf j}_{ij}(q) +\dots\ .
\end{eqnarray}
Calculations that retain only one-body terms in Eq.~(\ref{operators}) are typically called impulse approximation (IA)
computations.
In this paper, instead, we denote with IA those calculations that make use of only the leading-order (LO) one-body term
in the chiral expansion of the electromagnetic operator~\cite{Pastore08,Pastore09,Pastore11,Kolling09,Kolling11,Kolling12}, basically
excluding the relativistic one-body corrections.
These operators are the standard charge and current one-body operators obtained from the
non-relativistic reduction of the covariant nucleonic electromagnetic currents.
In this work, we use two-body currents derived from a chiral effective field theory
with pions and nucleons up to and including one-loop corrections~\cite{Pastore08,Pastore09,Pastore11,Kolling09,Kolling11,Kolling12}. 
Note that, contributions from two-body terms enter at next-to-leading order (NLO) and at N$^4$LO
in the chiral expansion of the current and charge operators, 
respectively. Thus, two-body terms are expected to be sizable 
in observables induced by the current operator and small in those induced by the charge operator. 

The finite size of the nucleon is accounted for by including 
the proton ($p$) and neutron ($n$) electric ($E$) and magnetic ($M$) form 
factors, $G^{p/n}_{E/M}(q^2)$. For example, the IA charge operator in the point-nucleon limit reads
\begin{equation}
\rho({\bs q})\!= \! \sum_i^Z \exp( {\bs q} \cdot {\bs r}_i)\,,
\end{equation}
where ${\bs r}_i$ is the coordinate of the $i$-th nucleon, and it becomes
\begin{equation}
\rho({\bs q})\!=\!\! \sum_i^A \! \exp({\bs q}\cdot {\bs r}_i)\!  \left [ \left( \frac{1+\tau^3_i}{2} \right)\! G^p_E(q^2)+ \left( \frac{1-\tau^3_i}{2} \right)\! G^n_E(q^2)\! \right] \,,
\end{equation}
 when the nucleonic form factors are included,
with $\tau^3_i$ being the third isospin component of the $i$-th nucleon.
$G^{p/n}_{E/M}(q^2)$ are typically represented by parameterizations of 
electron-scattering data, and here we will test the sensitivity of our results
to different nucleonic inputs. 

The nuclear electromagnetic form factors $F_E$ and $F_M$
can be regarded as distributions, and thus one can define the corresponding 
momenta at different orders in the $q^2$ expansion.
The 2$^{\rm nd}$ and 4$^{\rm th}$ electric (magnetic) moments can be 
derived from an expansion near momentum transfer $q^2\rightarrow 0$ of the charge 
(magnetic) form factor as 
\begin{equation}
\label{eq:ffs}
 F_{x}(q^2)= 1 -\frac{1}{3!}\langle R_x^2 \rangle q^2 +\frac{1}{5!}\langle R_x^4 \rangle q^4 + \dots  \ ,
\end{equation}
where
\begin{eqnarray}
 \langle R_x^2 \rangle &=& -\,6 \, \frac{\partial F_x(q^2)}{\partial q^2}\bigg|_{q=0} \ , \\
 \langle R_x^4 \rangle &=& 60 \, \frac{\partial^2 F_x(q^2)}{\partial^2 q^2}\bigg|_{q=0} \ , 
\label{eq:rmom}
\end{eqnarray}
with $x=E (M)$.
Given the calculated nuclear form factors $F_x(q^2)$ at small values of 
$q^2$, the $\bra R_x^{2,4} \ket$ are then obtained from a quadratic fit 
as indicated in Eq.~(\ref{eq:ffs}). From these, of course, follow
estimates of, e.g.,  the rms  charge radius $R_{E}=\sqrt{\bra R_{E}^2\ket}$, 
which is measured by Lamb shift experiments in muonic atoms using Eq.~(\ref{eq:E2s2p}).

The elastic component of $\delta_{\rm TPE}$ in Eq.~(\ref{eq:E2s2p}), 
namely $\delta_{\rm Zem}$, is directly proportional to the third Zemach 
moment~\cite{Ji13}, defined as
\begin{equation}
\label{eq:zemmom}
 \ev{R^3_{E}}_{(2)}  = \frac{48}{\pi}\int_0^\infty \frac{dq}{q^4} \left[ F_E^2(q^2) -1 + \frac{q^2 \bra R_{E}^2\ket }{3} \right]\,.
 \end{equation}

The Zemach radius (traditionally called the first Zemach moment) 
is a quantity of mixed electric and magnetic nature, 
defined as
\begin{equation}
\label{eq:rz}
\bra R_Z \ket  = -\frac{4}{\pi}\int_0^\infty \frac{dq}{q^2} \left[ F_E(q^2)\,\frac{F_M(q^2)}{F_M(0)} -1  \right] \, .
\end{equation}
It was first developed by Zemach in Ref.~\cite{Zemach_1956_PR} in the context of hyperfine 
splitting in hydrogen $S$-states, where the leading correction due to the proton's finite 
size was shown to be proportional to  $\ev{R_Z}$.
Consequently, $\ev{R_Z}$ of a spin-half nucleus can be experimentally determined, e.g., 
from the hyperfine splitting in its muonic hydrogen-like atom/ion, with precision that 
could rival determinations from electron scattering~\cite{Antognini13,Pohl_2016_LEAP}.

$R_E=\ev{R_E^2}^{\frac{1}{2}}$, $\ev{R^3_{E}}_{(2)}$, $\ev{R_E^4}$, $\ev{R_Z}$, $R_M=\ev{R_M^2}^{\frac{1}{2}}$
and $\mu$ are the observables we study in this paper. Since they are all essentially moments
of electromagnetic distributions, we refer to them cumulatively as ``electromagnetic moments''.

We would like to comment on the $q$ integration that enter in the above definitions. Clearly for a certain large value of $q$, denoted as $Q_{max}$, the form factors $F_{E/M}(q^2)$ 
are too small to contribute to the integrals in Eqs.~(\ref{eq:zemmom}) and (\ref{eq:rz}). Therefore from $q=Q_{max}$ up to 
$q=\infty$ the tail of the integrand is given  by the analytical expression in Eqs.~(\ref{eq:zemmom}) and (\ref{eq:rz}) where $F_{E/M}(q^2)$ are set~to~0. 
On the other hand, the integrands of the above equations are numerically unstable near $q=0$.
Therefore at $0\leqslant q \leqslant Q_{min}$, where $Q_{min}$ is a small value, they are replaced with their low-$q^2$ approximations 
\begin{eqnarray}
\label{eq:r30}
&& \lim_{q^2\to 0} \!\frac{1}{q^4}\!\! \left[ F_E^2(q^2)\!-\!1\!+\! \frac{q^2  R_{E}^2}{3} \right]
\!\!= \!  \frac{R^4_{E}}{36}+\frac{\ev{R^4_{E}}}{60} , \\
&& \lim_{q^2\to 0} \!\frac{1}{q^2} \!\!\left[ F_E(q^2)\,F_M(q^2) -1  \right]\!=\! - \frac{\langle R_{E}^2\rangle}{6}-\frac{\langle R_{M}^2\rangle}{6}    .
\label{eq:r0}
\end{eqnarray}

Using Eqs.~\eqref{eq:ffs}--\eqref{eq:r0} to calculate electromagnetic moments 
is hereafter referred to as the momentum-space numerical procedure and denoted with ``$q$-space''.

\subsection{Coordinate-space formulation}
The rms charge radius, as well as other even moments, can be readily obtained from  point-nucleon computations in coordinate-space.
In the IA and in the non-relativistic limit, the  2$^{\rm nd}$ and 4$^{\rm th}$ moments of the electric charge distribution can be obtained  as
\begin{eqnarray}
\label{eq:coor}
\langle R^2_{E} \rangle & =& \langle R^2_p \rangle  +  {r_p}^2+ \frac{N}{Z} r^2_n, \\
\nonumber
\ev{R^4_{E}}& =& \langle R^4_p \rangle + r^4_{p} + \frac{N}{Z} r^4_{n}  \!
+ \!  \frac{10}{3}\!\! \left( r^2_{p} \langle R^2_p\rangle + \frac{N}{Z} r^2_{n} \langle R^2_n\rangle \!\! \right)\,,
\end{eqnarray}
where, the point-proton mean-square radius is calculated as an expectation value on the ground-state wave-function
\begin{equation}
  \langle R^2_p \rangle = \langle \Psi_0 |\frac{1}{Z}\sum_i^Z r^2_i | \Psi_0 \rangle \,.
\end{equation}
Analogous
expressions exist for the point-neutron radius $ \langle R^2_n \rangle$ and for $ \langle R^4_p \rangle$. 
We perform our benchmark calculations with the
Kelly parameterization of the nucleon form factors~\cite{Kelly_2004}.
Accordingly, the 2$^{\rm nd}$ and 4$^{\rm th}$ moments of the intrinsic nucleon electric form factors are taken to be %, respectively,
$r^2_p=0.744(7)$ fm$^2$, $r^2_n= -0.112(3)$ fm$^2$, 
$ r^4_p =1.6(1)$ fm$^4$, and $r^4_n= -0.33(2)$ fm$^4$.

Using Eqs.~\eqref{eq:coor} to calculate 
$\langle R^2_{E} \rangle$ and $\ev{R^4_{E}}$ will be 
referred to as the coordinate-space numerical procedure and denoted with ``$r$-space''. 

\subsection{Mixed momentum \& coordinate-space formulation}
Given $F_E(q^2)$, one can obtain the charge density in coordinate-space, in the non-relativistic limit, as its Fourier transform
\begin{equation}
\label{eq:rho-F}
\rho_{E} ({\bs r}) = \int d^3 q~ F_E(q^2) ~e^{-i \bs{q}\cdot \bs{r}}~.
\end{equation}
The $n$-th electric Zemach moment is defined as
\begin{equation}
\label{eq:Zem_exact_QR}
\bra R^n_{E} \ket_{(2)} = \int d^3 {r} \int d^3{r'}~
\rho_{E} (r')~\rho_{E} (|\bs{r'}-\bs{r}|)~r^n\,.
\end{equation}
By inserting Eq.~\eqref{eq:rho-F} into Eq.~(\ref{eq:Zem_exact_QR})
one obtains
\begin{equation}
\label{eq:Zem_exact_QR2}
\bra R^n_{E} \ket_{(2)} = 
\int_0^{\infty} dr\,r^{n+1}
\left[
\frac{2}{\pi} \int_0^\infty dq \, q F^{2}_{E}(q^2) \sin(qr)\right]\,,
\end{equation}
which contains integrals on both $q$ and $r$. 
In the above we used explicitly only the contribution from the ``spherical'' 
part of the charge distribution, which is an approximation for the deuteron, but is  exact for $A=$ 3 and 4. 
This algorithm was found to be very robust, 
and does not suffer the numerical uncertainty associated with the regularization of Eq.~\eqref{eq:zemmom} at q=0.

Obviously, $F^{2}_{E}(q^2)$ in Eq.~\eqref{eq:Zem_exact_QR2} may be replaced with either 
$F_{E}(q^2)\cdot F_{M}(q^2)$ or $F^{2}_{M}(q^2)$, leading to the calculation of other moments, 
such as, e.g., $\ev{R_Z}$ of Eq.~\eqref{eq:rz}.
Additionally, there exist relations between various Zemach and regular moments, 
e.g.,
\begin{eqnarray}
\label{eq:R2_from_Zem}
2 \bra R^2_{E}\ket &=& \ev{R^2_{E}}_{(2)}\,, \\
\label{eq:R4_from_Zem}
2 \ev{R^4_E} &=& \ev{R^4_E}_{(2)} - \frac{10}{3} \bra R^2_{E}\ket^2\,, 
\end{eqnarray}
which enable the consistent calculation of essentially all the regular and Zemach moments, and particularly all the observables targeted here, via this procedure.

Using Eqs.~\eqref{eq:Zem_exact_QR2}--\eqref{eq:R4_from_Zem} to
 calculate electromagnetic moments 
will be referred to as the mixed-space numerical procedure, which we label with ``$qr$-space''. 

\subsection{Numerical Procedures: Comparison}
We apply the numerical procedures detailed above to study the $^3$He electric moments. 
In particular, the $r$- and $qr$-space procedures are used in combination with EIHH
few-body computational method, while the $q$-space procedure is implemented within the 
HH-p method. 
\begin{center}
  \begin{table}[hbt]
\caption{Calculations of the $^3$He charge radius, $R_E$, 3$^{\rm rd}$ Zemach moment, $\bra R^3_E\ket_{(2)}$, 
and 4$^{\rm th}$  electric moment, $\bra R^4_E\ket$  in IA, based 
on the  AV18+UIX nuclear interaction, and Kelly parameterization for the nucleonic form factors.
The errors in parenthesis account for the computational error associated with the few-body method
(either EIHH or HH-p) and the error from the numerical procedure ($r$-, $qr$-, or $q$-space) added in quadrature.}
\begin{tabular}{l l }																
\hline	
$R_E$ [fm]					&$\qquad 	^3$He \\%		&$\qquad\quad	^3$H		\\
$r$-space (EIHH)	&$\ $	1.953(2)			       \\%    &$\quad$	1.7399(0)(2)(7)		\\
$qr$-space (EIHH)	&$\ $	1.953(2)			      \\%&$\quad$ 	1.7398(0)(9)(5)		\\
$q$-space (HH-p) 	&$\ $	1.953(1)			\\%	&$\quad$ 	1.7399(-)(5)(-)		\\
\hline		
$\bra R^3_E\ket_{(2)}$ [fm$^3$]	&$\qquad 	^3$He		\\%&$\qquad\quad	^3$H		\\
$r$-space (EIHH)         &$\ $   --			       \\%    &$\quad$	1.7399(0)(2)(7)		\\
$qr$-space (EIHH)	 &$\ $	27.65(10)			\\%&$\quad$	19.30(0)(3)(2)		\\
$q$-space (HH-p)	 &$\ $	27.56(20)			\\%&$\quad$ 	19.30(2)(5)(7)		\\
\hline															
$\bra R^4_E\ket$ [fm$^4$]		&$\qquad 	^3$He		\\%&$\qquad\quad	^3$H		\\
$r$-space (EIHH)	&$\ $	33.88(52)			\\%&$\quad$	19.69(0)(1)(45)		\\
$qr$-space (EIHH)	&$\ $	33.79(24)			\\%&$\quad$ 	19.95(1)(5)(04)		\\
$q$-space (HH-p)	&$\ $	32.5(1.3)			\\%&$\quad$ 	19.30(2)(5)(7)		\\
\hline										
\end{tabular}
\label{tab:procedure}
\end{table}
\end{center}

In Table~\ref{tab:procedure}, we compare results 
obtained using the AV18 two-body (NN) nuclear force~\cite{AV18} complemented by 
the Urbana IX (UIX) three-body (3N) force~\cite{Pudliner95}---which we denote with AV18+UIX,
and the Kelly nucleonic form factors~\cite{Kelly_2004}. 
The values in brackets are estimates of the computational uncertainties corresponding to 
the numerical procedure and the computational method added in quadrature. 
The various procedures produce consistent  results within 
uncertainties. The latter are typically of the order of 0.1$\%$  for $R_E$ and 0.4-0.7$\%$ for the third 
Zemach  moment.
For the  fourth moment, instead,  the $q$-space procedure, while being statistically in agreement with the
other estimates, is affected by a larger ($\sim 4\%$) uncertainty, while the $r$- and $qr$-procedure lead to an uncertainty of about 1$\%$.
Here we remark that the uncertainty from the $q$-space
extrapolation could potentially affect also experimental extractions of higher 
moments. Overall, we find that the mixed ($qr$-space) procedure is more robust 
and allows for higher precision, without the need to investigate the quality 
of the fitting and regulating procedures corresponding to 
Eqs.~\eqref{eq:ffs}--\eqref{eq:r0}.%~\\

  \begin{center}
    \begin{table*}[t]
  \caption {Deuteron benchmark in IA: calculations with the harmonic 
  oscillator basis (HO) or the Numerov algorithm based on the AV18 potential, 
  in the point-nucleon limit, i.e., without form factors (w/o FF), or 
  with nucleon finite sizes parameterized by the Kelly form factors 
  (w FF). Experimental data are shown in the last row.
   }
      \label{deuteron}
  \begin{tabular}{l l l l l l l}
& \multicolumn{1}{c}{$R_{E}$} 	
& \multicolumn{1}{c}{$\ev{R^3_{E}}_{(2)}$}	
& \multicolumn{1}{c}{$\ev{R^4_{E}}$} 
& \multicolumn{1}{c}{$\ev{R_{\rm Z}}$}	
& \multicolumn{1}{c}{$R_{M}$}	
& \multicolumn{1}{c}{$\mu$}	\\
Method	& \multicolumn{1}{c}{[fm]}			 	& \multicolumn{1}{c}{[fm$^3$]}							& \multicolumn{1}{c}{[fm$^4$]}  & \multicolumn{1}{c}{[fm]}	& \multicolumn{1}{c}{[fm]}	& \multicolumn{1}{c}{[$\mu_N$]}	\\
\hline	 
HO (w/o FF)		&$\quad\;$ \rmsPP 	&$\;$\thirdZemPP		&$\;$\rfourthPP	&$\quad\quad$\rzPP	&$\quad\quad$\rmagPP		&\quad\quad\quad\quad${0.84699(1)}$ \\
Numerov (w/o FF)	&$\quad\;$ 1.9674(1) 	&$\;$31.83(1)		&$\;$55.376(1)	&$\quad\quad$2.3795(1)	&$\quad\quad$1.9405(1)		&\quad\quad\quad\quad${0.84699(1)}$ \\
HO (w FF)		&$\quad\;\;$\rmsFF	&$\;$\thirdZemFF		&$\;$\rfourthFF	&$\quad\quad$\rzFF 	&$\quad\quad$\rmagFF	&\quad\quad\quad\quad${0.84699(1)}$ \\
Numerov (w FF)     	&$\quad\;$ 2.1218(1) 	&$\;$38.33(1)		&$\;$64.814(1) &$\quad\quad$2.595(3)	&$\quad\quad$2.0664(1)		&\quad\quad\quad\quad${0.84699(1)}$ \\
\hline 
\rule{0pt}{4ex}
Exp.	 			&$\quad\;\;$2.1413(25) \cite{CODATA_2014}	&$\;$n.a   &$\;$n.a &$\;\;\;$2.593(16)  \cite{Friar_2004}  & $\;\;\;$1.90(14) \cite{Afanasev_1998}	&\qquad 0.8574382311(48)~\cite{CODATA_2014}   \\
                                &$\quad\;\;$2.1256(8) ~\cite{Pohl_2016_Science} 
\end{tabular}
\end{table*}
  \end{center}

%%%%%%%%%%%%%%%%%%%%%%%%%%%%%%%%%%%%%%%%%%%%%%%%%
\section {Results}
\label{sec:Results}
In this section we present results for $A=2,3$ and $4$ nuclei. 
Following the investigation outlined above,  we will show EIHH
results obtained using the $qr$-procedure %(unless otherwise specified) 
and HO results for the deuteron obtained using the $r$-space procedure,  
which involved the least approximation in this case.
Numerov results and HH-p results use the $q$-space procedure, 
while quantum Monte Carlo results use the $qr$-space procedure.

%%%%%%%%%%%%%%%
\subsection {Benchmark in impulse approximation}
\label{sec:ben}

%\begin{widetext}
\begin{center}
\begin{table*}%[hb]
\caption{ $^3$He electromagnetic moments calculated in IA with several ab initio methods using the AV18+UIX 
nuclear Hamiltonian. Experimental values are from Ref.~\cite{Sick2014,PURCELL20101}.
Errors in parenthesis are from the computational method and the numerical procedure applied 
to extract the  moments. See text for explanations.}
 \begin{tabular}{ l l l l l l l}
&
\multicolumn{1}{c}{$R_{E}$} 
	& \multicolumn{1}{c}{$\ev{R^3_{E}}_{(2)}$}	
		& \multicolumn{1}{c}{$\ev{R^4_{E}}$} 
			& \multicolumn{1}{c}{$\bra R_{\rm Z}\ket $}	
				& \multicolumn{1}{c}{ $R_{M}$}	
					& \multicolumn{1}{c}{$\mu$}	\\
Method	
	& \multicolumn{1}{c}{[fm]}			 	
		& \multicolumn{1}{c}{[fm$^3$]}							
			& \multicolumn{1}{c}{[fm$^4$]}  
				& \multicolumn{1}{c}{[fm]}	
					& \multicolumn{1}{c}{[fm]}	
						& \multicolumn{1}{c}{[$\mu_N$]}	\\
\hline	 
\rule{0pt}{3ex}
VMC	    	&$\;$1.956(1)		&$\;$27.8(1) 	&$\;$33.5(1)		&$\;$2.58(1)		&$\;$2.000(1)		&${-1.774}(1)$ \\
GFMC    	&$\;$1.954(3)		&$\;$27.7(2) 	&$\;$33.7(4)		&$\;$2.60(1)  		&$\;$1.989(8)		&${-1.747(2)}$ \\[5pt]
HH-p    	&$\;$1.953(1)		&$\;$27.56(20) 	&$\;$32.5(1.3)	&$\;$2.598(1)		&$\;$2.103(1)		&${-1.757(1)}$ \\[5pt]
EIHH    	&$\;$1.953(2) 	        &$\;$27.65(10)	&$\;$33.8(2)		& $\qquad$-		&$\qquad$-		&$-1.758(1)$	  \\
\hline 
\rule{0pt}{4ex}
Exp.		&$\;$1.973(14)	&$\;$28.15(70)		&$\;$32.9(1.60)	&$\;$2.528(16)	&$\;$1.976(47)		&$\;{-2.127}$ 
\end{tabular}
\label{tab:benhe3}
\end{table*}
\end{center}

%\begin{widetext}
\begin{center}
\begin{table*}%[htb]
\caption{Same as Table~\ref{tab:benhe3} but for $^3$H. Experimental values are from Refs.~\cite{ANGELI201369,PURCELL20101}.}
 \begin{tabular}{l  l l l l l l}
& \multicolumn{1}{c}{$R_{E}$} 	
& \multicolumn{1}{c}{$\ev{R^3_{E}}_{(2)}$}	
& \multicolumn{1}{c}{$\ev{R^4_{E}}$} 
   & \multicolumn{1}{c}{$\bra R_{\rm Z} \ket$}	
& \multicolumn{1}{c}{$R_{M}$}	
& \multicolumn{1}{c}{$\mu$}	\\
Method	
& \multicolumn{1}{c}{[fm]}			 	
& \multicolumn{1}{c}{[fm$^3$]}							
& \multicolumn{1}{c}{[fm$^4$]}  
& \multicolumn{1}{c}{[fm]}	
& \multicolumn{1}{c}{[fm]}	
& \multicolumn{1}{c}{[$\mu_N$]}	\\
\hline	 
\rule{0pt}{3ex}
VMC     		&$\;$1.765(1)	&$\;$20.2(1)		&$\;$21.1(1)		&$\;$2.37(1)	&$\;$1.898(1)		&2.588(1)	\\
GFMC     		&$\;$1.747(2)	&$\;$19.6(1)		&$\;$20.0(2)		&$\;$2.35(1)  	&$\;$1.899(7)		&2.555(2)\\[5pt]
HH-p     		&$\;$1.745(1)	&$\;$19.34(13)	&$\;$19.0(4)		&$\;$2.355(1)	&$\;$1.922(1)		&2.579(1) \\[5pt]
EIHH     		&$\;$1.740(1)	&$\;$19.30(4)	&$\;$19.95(6)		& $\qquad$-	&$\qquad$-		&$2.572(1)$	  \\
\hline 
\rule{0pt}{4ex}
Exp.		 	&$\;$1.759(36)	&$\qquad$-			&$\qquad$-			&$\qquad$-		&$\;$1.840(181)		&2.979
\end{tabular}
\label{tab:benh3}
\end{table*}
\end{center}
%\end{widetext}
%\end{widetext}

First, we benchmark electromagnetic moments of $^2$H, $^3$H, $^3$He and $^4$He 
calculated in IA, which include nucleon form factors from the 
Kelly parameterization~\cite{Kelly_2004}. 
Ground-state wave-functions were obtained from the AV18 two-body nuclear
interaction for the deuteron, and the AV18+UIX nuclear Hamiltonian
for $A=3,4$ nuclei. We will neglect isospin 
symmetry breaking (ISB) effects in $A=2$ and $3$, which were found 
to be small~\cite{NND16}.

In Table ~\ref{deuteron}, we show results for the deuteron 
calculated expanding on the harmonic oscillator basis or 
using the Numerov algorithm. We  show results both in the point-nucleon 
limit, i.e., without form factors (w/o FF), and when the nucleon finite 
sizes are included via the Kelly parameterization (w FF).
The two numerical methods are in  perfect agreement 
with each other for all the observables except the Zemach radius, 
third Zemach and fourth charge moments. The latter are more 
sensitive to the numerical procedure but the differences are 
not significant ($\sim 0.1\%$). 
The inclusion of finite size effects via the 
nucleon form factors improves the agreement with 
experiment for all the observables expect for the magnetic radius. 
For magnetic properties it is known that the addition of two-body currents %, namely going beyond the IA, 
is required to explain the experimental data~\cite{Schiavilla:2018udt}. 
For  $\mu$, calculations with or without form factors are the same in IA, since, at leading order,
finite size effects are proportional to $q$, thus they are suppressed in the limit $q\rightarrow 0$.
%%%%%%%%%%%%%%%
%\paragraph {A=3}
%\label{sec:A3}

Next, we benchmark  $^3$He and $^3$H  electromagnetic moments in IA, 
where we solve the Schr\"odinger equation with the VMC, GFMC, HH-p, and EIHH computational methods.
The results, which include nucleon finite sizes via the Kelly parameterization and are 
presented in  Tables~\ref{tab:benhe3} and \ref{tab:benh3} with computational uncertainties.
Specifically, these uncertainties are a quadrature sum of 
the uncertainties from the numerical procedure described in the previous paragraph and
those coming from the few-body method, e.g., due to truncation of the model-space for 
basis expansion methods or statistical uncertainties for Monte Carlo methods.
When using the $q$-space procedure, the former were typically larger than the latter.

Comparing the results from the different few-body methods, we observe that they are consistent and in  
 agreement with each other for $^3$He, while for $^3$H the $R_{E}$ and $\ev{R^4_{E}}$ values 
obtained with the EIHH are slightly smaller than with the other methods.
Although this difference is not significant,
it is found to be consistent with available literature, where, e.g., 
HH calculations with AV18+UIX reported in Ref.~\cite{Kievsky_2008_JPG} 
produce $\ev{R^2_p}$ of \nuc{3}{H} that is smaller by $\sim 1$\% than GFMC calculations
of Refs.~\cite{Pieper_2001_PRC}.
The small  differences on $R_M$ and $\mu$  are possibly due to the fact that
magnetic observables probe also the spin degrees of freedom and thus are more sensitive to details 
in the wave functions.

Interestingly, one observes that in IA the electromagnetic moments (magnetic moments) 
are overestimated (underestimated) with respect to the experiment. This is due 
to the missing contributions from relativistic corrections and two-body currents, that will 
be discussed in the next section.

Finally, in Table~\ref{He4}, we present the  \nuc{4}{He} electric moments in IA. 
 For this nucleus, we explore the effect of isospin symmetry breaking (ISB) within the EIHH method.
 We denote this last set numerical values with EIHH-ISB. As one can see, ISB 
 effects are rather small (between $0.1\%$ and $0.6\%$).

 Comparing the various methods, one sees that VMC and GFMC are very close to 
 each other for $^4$He, more so than for the three-body nuclei.
 The EIHH values are consistently smaller, and the  ISB terms
  systematically enhance the electric radii.
 Compared to experiment, theoretical calculations in IA underestimate
 the measurements by a few percent, similarly to what it is found in the $A=3$ nuclei.

%%%%%%%%%%%%%%%
%\subsubsection {A=4}
%\paragraph {A=4}
%\label{sec:A4}

%\vspace{5cm}
 %%%%%%%%%%%%%%%%%%%%%%%%%%%%%%%%%%%%%%%%%%%%%%%He4
 \begin{center}
   \begin{table}%[hbt]
 \caption{$^4$He electric moments calculated in IA with several ab initio methods using the AV18+UIX 
nuclear Hamiltonian. Experimental values are from Ref.~\cite{Sick2014}. Uncertainties in parenthesis are from 
the computational method and the numerical procedure applied to extract the moments. See text for 
explanations.}
  \begin{tabular}{l  l l l l l l}
 %\multicolumn{2}{l}{}					
 & \multicolumn{1}{c}{$R_{E}$} 	
 & \multicolumn{1}{c}{$\ev{R^3_{E}}_{(2)}$}	
 & \multicolumn{1}{c}{$\ev{R^4_{E}}$} \\
 Method	
 & \multicolumn{1}{c}{[fm]}			 	
 & \multicolumn{1}{c}{[fm$^3$]}							
 & \multicolumn{1}{c}{[fm$^4$]}  \\
 \hline	 
 \rule{0pt}{3ex}
 VMC				&$\;$1.649(1)		&$\;$16.0(1)		&$\;$14.1(1)	\\
 GFMC			&$\;$1.648(2)		&$\;$16.0(1)		&$\;$14.1(1)	\\[5pt]
 EIHH	    	   	&$\;$1.638(2)		&$\;$15.6(2)		&$\;$13.6(2)	\\
 EIHH-ISB			&$\;$1.640(2)		&$\;$15.7(2)		&$\;$13.7(2)	\\ 		
 \hline 
 \rule{0pt}{4ex}
 Exp.				&$\;$1.681(4)		&$\;$16.73(10)	&$\;$14.35(11)
 \end{tabular}
 \label{He4}
 \end{table}
 \end{center}

% 
% 
%%%%%%%%%%%%%%%
\subsection {Two-body currents and relativistic corrections}
\label{sec:MEC}
The results reported in the previous section are obtained using 
charge and current operators in IA.
Here, we study the contributions generated by one-body 
relativistic corrections (RC), and two-body components in the 
electromagnetic currents. We use electromagnetic currents
derived from chiral effective field theory in Refs.~\cite{Pastore08,Pastore09,Pastore11,Piarulli12,Kolling09,Kolling11}.
In particular, we adopt the implementation in the HH-p scheme described in  
Ref.~\cite{Piarulli12}.
%At this time, two-body currents have been implemented only within the 
%HH-p method, therefore, we will size the effects of RC and two-body currents
%in $A=3$ nuclei within this framework.
We emphasize that the calculations we present are hybrid, meaning that chiral 
currents are used in combination with wave functions obtained from the
AV18+UIX nuclear interactions. Intrinsic to this approach is a mismatch 
between the short-range dynamics used to correlate nucleons in pairs 
and that implemented in the two-body current operators. Additional uncertainties
arising from this procedure will be discussed briefly in Section~\ref{sub:models}. 
Calculations of electromagnetic observables in $A=2$ and $3$ system based
on both chiral currents and interactions have been recently performed in 
Ref.~\cite{Schiavilla:2018udt}, and detailed studies of electromagnetic 
moments within a chiral formulation will be possible in the near future.

% RC+MEC TABLES
\begin{widetext} 
\begin{center}
  \begin{table*}%[htb]
    \caption{Electromagnetic moments for $^3$He calculated within the HH-p method. 
    Beyond the IA, relativistic corrections (RC) are included, as well as two-body 
    currents~\cite{Pastore08,Pastore09,Pastore11,Piarulli12} which are added
    to the IA+RC results and reported in column labeled with TOT. Experimental values are 
    from Refs.~\cite{Sick2014,PURCELL20101}. Errors reported in the second parenthesis
    of the final results (TOT) account for uncertainties due to the truncation in the 
    chiral expansion. See text for explanations.  
    }
 \begin{tabular}{l  l l l l l l}
& \multicolumn{1}{c}{$R_{E}$} 	
& \multicolumn{1}{c}{$\ev{R^3_{E}}_{(2)}$}	
& \multicolumn{1}{c}{$\ev{R^4_{E}}$} 
& \multicolumn{1}{c}{$R_{\rm Z}$}	
& \multicolumn{1}{c}{$R_{M}$}	
& \multicolumn{1}{c}{$\mu$~~~~}	\\
Method	
& \multicolumn{1}{c}{[fm]}			 	
& \multicolumn{1}{c}{[fm$^3$]}							
& \multicolumn{1}{c}{[fm$^4$]}  
& \multicolumn{1}{c}{[fm]}	
& \multicolumn{1}{c}{[fm]}	
& \multicolumn{1}{c}{[$\mu_N$]}~~	\\
\hline	 

IA     	  &$\;$1.953(1)	        &$\;$ 27.56(20) 	&$\;$32.5(1.3)		&$\;$2.598(1)    &$\;$2.103(1)	& -1.757(1)	\\
IA+RC     &$\;$1.975(1)      	&$\;$ 28.44(20) 	&$\;$33.6(1.3)		&$\;$2.621(1)    &$\;$2.116(1)	& -1.737(1)         \\
TOT       &$\;$1.979(1)(10)	&$\;$ 28.58(66)(13)     &$\;$33.8(1.5)(2)	&$\;$2.539(3)(19) &$\;$1.991(1)(31)	& -2.093(1)(55)   \\
\hline 
%\rule{0pt}{4ex}
Exp.			&$\;$1.973(14)	&$\;$~28.15(70)	&$\;$32.9(1.60)	&$\;$2.528(16)	&$\;$1.976(47)		&-2.127 
\end{tabular}
\label{tab:he3mec}
\end{table*}
\end{center}
%\end{widetext}

%\begin{widetext}
%\begin{center}
\begin{table*}%[htb]
\caption{Same as Table~\ref{tab:he3mec} but for $^3$H. Experimental values are from~\cite{ANGELI201369,PURCELL20101}. }
 \begin{tabular}{l  l l l l l l}
& \multicolumn{1}{c}{$R_{E}$} 	
& \multicolumn{1}{c}{$\ev{R^3_{E}}_{(2)}$}	
& \multicolumn{1}{c}{$\ev{R^4_{E}}$} 
& \multicolumn{1}{c}{$R_{\rm Z}$}	
& \multicolumn{1}{c}{$R_{M}$}	
& \multicolumn{1}{c}{$\mu$}	\\
Method	
& \multicolumn{1}{c}{[fm]}			 	
& \multicolumn{1}{c}{[fm$^3$]}							
& \multicolumn{1}{c}{[fm$^4$]}  
& \multicolumn{1}{c}{[fm]}	
& \multicolumn{1}{c}{[fm]}	
& \multicolumn{1}{c}{[$\mu_N$]}	\\
\hline	 
IA     		&$\;$1.745(1)	&$\;$ 19.34(13) 	&$\;$19.0(4)	&$\;$ 2.355(1)     &$\;$1.922(1)      	& 2.579(1)	\\
IA+RC            &$\;$1.716(1)	&$\;$ 18.35(13) 	&$\;$17.6(4)	&$\;$ 2.347(1)     &$\;$1.936(1)      	& 2.542(1)      \\
TOT     	&$\;$1.726(2)(9)&$\;$ 18.61(37)(8)	&$\;$17.6(1)(1)	&$\;$ 2.295(3)(24)  &$\;$1.850(1)(30)	& 2.955(1)(74)       \\
\hline 
\rule{0pt}{4ex}
Exp.	        	&$\;$1.759(36)	&$\qquad$-     &$\qquad$-	&$\qquad$-	&$\;$1.840(181)		&2.979
\end{tabular}
\label{tab:h3mec}
\end{table*}
%\end{center}
\end{widetext}

%Because two-body operators have very little contribution in the isoscalar 
%deuteron (in the charge operator they appear at 
%next-to-next-to-next-to-leading order and in the current operator they appear at at next-to-leading order, but they are 
%purely isovector),  we focus the rest of the discussion on the $A=3$ nuclei.

In Tables~\ref{tab:he3mec} and~\ref{tab:h3mec}, besides the calculations in IA, 
we show results obtained by adding relativistic corrections---column labeled with ``IA+RC'', and 
final results that include also two-body currents---column labeled with ``TOT''. 
We find that RC contributions are of the order of $1\%$ in both the charge and 
magnetic radii of the trinucleon systems
%~\footnote{
%We remark that the numbers reported in this work differ from those
%given in Ref.~\cite{Piarulli12}. These discrepancies are due to different 
%interpolation algorithms implemented by the two calculations and will 
%be discussed in more detail in Ref.~\cite{errata}.
%}, 
while they provide a 3$\%$--7$\%$ correction to the third Zemach and fourth electric
moments of $^3$He and $^3$H. The addition of RC significantly improves the 
comparison with experiments for the electric moments. As expected, 
the effect of two-body operators is very small for these observables while
it is sizable for the magnetic radii ($\sim 6\%$) and magnetic moments 
($\sim 15\%$), bringing the theoretical results in agreement with 
the experimental data.

In Tables~\ref{tab:he3mec} and~\ref{tab:h3mec} 
the error reported in the first bracket include the 
``$q$-space'' uncertainty---mostly coming from the 
fitting procedure described in the previous section---and 
uncertainties due to the few-body method, added in quadrature. 
These are the only uncertainties we report in the ``IA'' and ``IA+RC'' 
calculations, to be consistent with the benchmark results presented 
in the previous section. 
 %Overall, such uncertainties are negligible for the standard moments and of the order of $2\%$ for the third Zemach moments.
The uncertainty shown in the second bracket---which we provide
only for the final results labeled with ``TOT''---is an estimate of 
the error due to the truncation in the chiral expansion of the currents, 
here included up to one-loop. To estimate this theoretical 
uncertainty, we use the algorithm developed by Epelbaum {\it et al.} 
in Ref.~\cite{Epelbaum2015}.
The algorithm has in fact been applied to calculate the uncertainty given in
the second brackets of all moments except for $\ev{R_E^{(3)}}$ and $R_Z$. These observable 
are defined in  Eqs.~(\ref{eq:zemmom}) and~(\ref{eq:rz}). In particular, 
$R_Z$ involves a convolution of both the electric and magnetic form factors, 
induced by the charge and current operators, respectively. The theoretical error 
from the truncation in the chiral expansion, in these cases, is inferred from those
associated with $R_E$, $R_M$, and $\ev{R_E^4}$. For these observables, we utilize 
their expressions in the low-$q$ regime given in Eqs.~(\ref{eq:r30}) 
and~(\ref{eq:r0}), and obtain their theoretical errors by propagating 
those associated with $R_E$, $R_M$, and $\ev{R_E^4}$.  

The chiral truncation uncertainties are of the order of or less than
$\sim 1\%$ for the charge-radii, third Zemach and fourth moments, while they are 
of the order of $1\%$--2.5$\%$ for the magnetic radii, magnetic moments and first 
Zemach moments. In the case of the fourth moment
of $^3$He the $q$-space uncertainty ($\sim 4\%$) is comparable to and even 
larger than the chiral uncertainty. 
% The uncertainty from the $q$-space extrapolation
% could potentially affect also experimental extractions of the fourth moment.

\begin{figure}%[htb]
\centering
\includegraphics[width=0.5\textwidth]{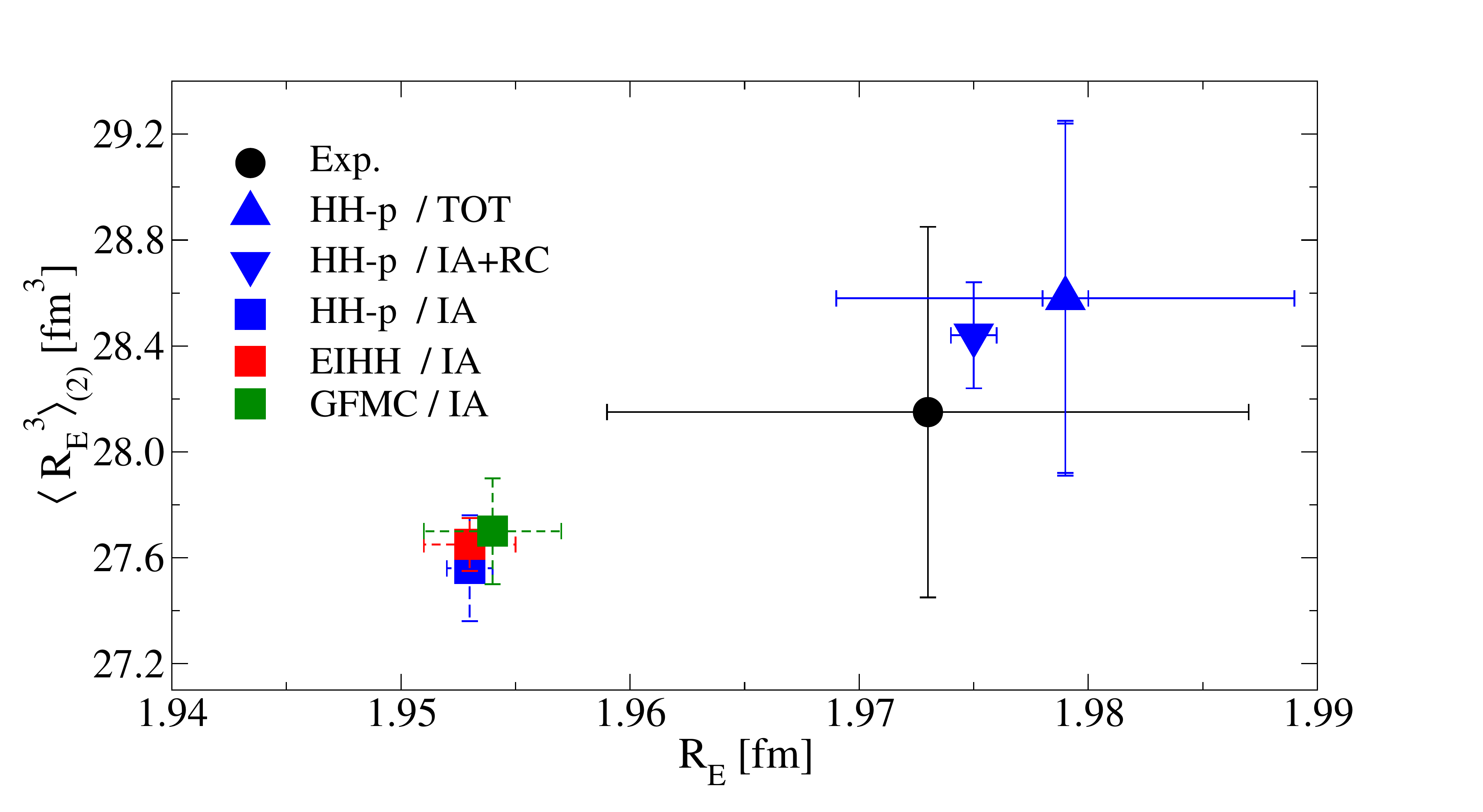}
\caption{(Color online) Third Zemach moment vs charge radius of $^3$He: 
different calculations are compared with experimental data from Ref.~\cite{Sick2014}. 
The calculations are based on the AV18+UIX nuclear potential and Kelly
form factors. Results with the charge operator in IA are labeled with IA, 
and those with the addition of RC (and two-body currents) are labeled with IA+RC 
(TOT). IA and IA+RC results are shown with uncertainties from the computational
methods alone, while the calculations labeled with ``TOT''
include also the chiral truncation error. See text for details.}
\label{Fig_Methods}
\end{figure}

\begin{figure}%[htb]
	\includegraphics[width=0.5\textwidth]{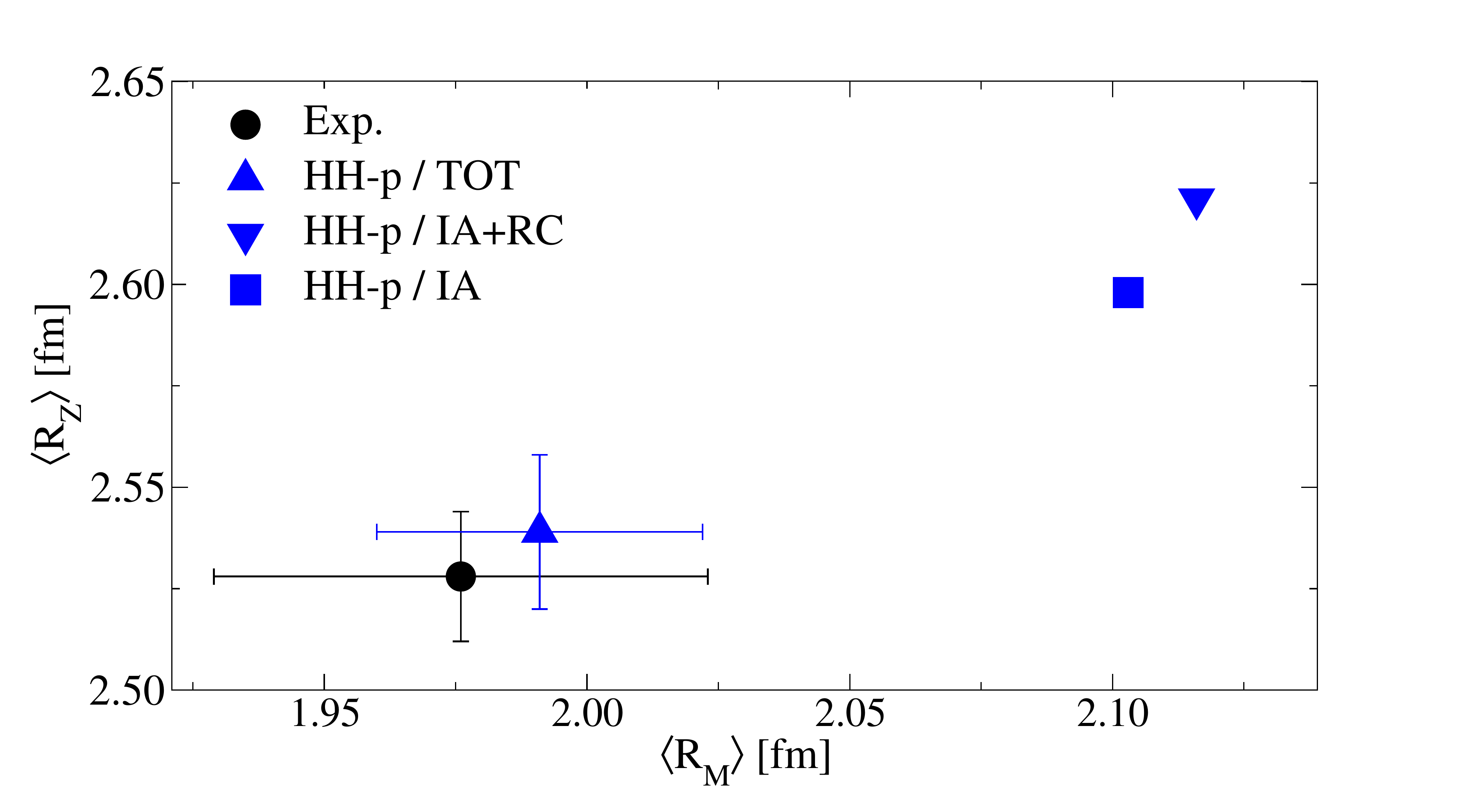} 
	\caption{(Color online) First Zemach moment  vs magnetic radius of $^3$He: 
	various calculations are compared with experimental data from Ref.~\cite{Sick2014}.
	Results with the charge operator in IA are labeled with IA, 
and those with the addition of RC (and two-body currents) are labeled with IA+RC 
(TOT). IA and IA+RC results are shown with uncertainties from the computational
methods alone, while the calculations labeled with ``TOT''
include also the chiral truncation error.
When not visible, error bars are included in the size of the symbols.  See text for details.
	}
	\label{figure_3He_mag}
\end{figure}

%
We combine the $^3$He results given in Tables~\ref{tab:benhe3} \& \ref{tab:he3mec} in 
Figs.~\ref{Fig_Methods} and \ref{figure_3He_mag}. In Fig.~\ref{Fig_Methods}, we plot 
the third Zemach moment versus  $R_E$ and compare  calculations using different numerical methods to  
data from electron scattering experiments~\cite{Sick2014}. First, one observes that 
the IA calculations obtained from different methods (EIHH, HH-p and GFMC) all agree 
within computational error bars---{\it albeit} they underestimate the experimental 
results--- thus demonstrating that the numerical uncertainties from the choice of  
the few-body techniques and numerical integration procedures are negligibly small.
Therefore, for these light nuclei, any of these few-body methods or numerical 
procedure may be used to further analyze the dependence on dynamical inputs, 
i.e., nucleonic form factors, nuclear Hamiltonians and two-body currents.
Second,  as expected, we observe  a strong correlation between the two plotted 
observables: they are roughly linearly correlated.
After the inclusion of RC and two-body electromagnetic currents, 
which combined together provide a $3\%-5\%$ contribution, the calculated 
observables are in very good agreement with the experimental values. 
The theoretical uncertainty of the results labeled with ``TOT'' includes
the chiral truncation error, which is summed in quadrature together
with the few-body and numerical
procedure uncertainties. 
In essence, the final results (``TOT'') account for a more complete uncertainty 
budget---as opposed to the other points shown in Fig.~\ref{Fig_Methods}--, 
which amounts to $\sim 0.5\%$ ($\sim 2\%$) for $R_E$ (third Zemach moment), 
comparable to the experimental uncertainty. 

In Fig.~\ref{figure_3He_mag}, we show the magnetic observables, 
namely $R_Z$ versus $R_M$, calculated with the HH-p method. 
Also in this case, we observe a correlation between the 
two observables. In particular, the IA over-estimates experiment, 
and RC have a smaller effect ($\sim 3\%$) than the two-body currents ($\sim 6\%$). Also in this case, 
once RC and two-body currents are included, theoretical results agree nicely with experiment. 
When the chiral truncation error is accounted for (again only in the point labeled 
with ``TOT''), theory and experiment have comparable uncertainties.

 %%%%%%%%%%%%%%%
 \subsection{Nuclear and nucleon models}
 \label{sub:models}
At this point, we briefly address the dependency on variations in  
two inputs that were kept fixed until now, namely, the nuclear 
interaction and the nucleonic form factors.
The effect due to a possible variance in each of these inputs may be 
considered as an additional source of uncertainty.  
To this end, we study the $\ev{R^3_E}_{(2)}$ and $R_E$ of $^3$He using the 
EIHH few-body method and the charge operator in IA. 

\begin{center}
   \begin{table}[hbt]
 \caption{$^3$He electric moments in IA calculated with
 the EIHH method using i) either the AV18+UIX or the $\chi$EFT nuclear Hamiltonian,  and
ii) different parameterizations of the nucleonic
form factor $G_E$. Quoted uncertainties include only the 
method and procedure error bars.  Experimental values 
are from Ref.~\cite{Sick2014,PURCELL20101}. See text for details.}
  \begin{tabular}{l  l l l l l }
 %\multicolumn{2}{l}{}					
 & \multicolumn{1}{c}{$R_{E}$} 	
 & \multicolumn{1}{c}{$\ev{R^3_{E}}_{(2)}$}\\
 Potential/$G_E$	
 & \multicolumn{1}{c}{[fm]}			 	
 & \multicolumn{1}{c}{[fm$^3$]}	\\						
 \hline	 
 %\rule{0pt}{3ex}
  $\chi$EFT/CODATA	       	&$\;$1.976(2)		&$\;$28.33(14)	\\	
  $\chi$EFT/Kelly	   	&$\;$1.971(2)		&$\;$28.20(14)	\\	
  $\chi$EFT/CREMA	       	&$\;$1.961(2)		&$\;$27.72(14)	\\	
  \hline 
 AV18+UIX/CODATA	    	&$\;$1.958(2)		&$\;$27.77(10)	\\	
 AV18+UIX/Kelly	    	   	&$\;$1.953(2)		&$\;$27.65(10)	\\	
 AV18+UIX/CREMA	    	   	&$\;$1.943(2)		&$\;$27.17(9)	\\	
\hline
 \rule{0pt}{4ex}
 Exp.			&$\;$1.973(14)	      &$\;$~28.15(70)	\\
 \end{tabular}
 \label{Table_new}
 \end{table}
 \end{center}

In order to provide a rough estimate 
of the overall nuclear model dependency, 
% assign an additional uncertainty intrinsic to the hybrid
%approach we implemented here, 
%Because a thorough study of systematical and statistical uncertainties
%as performed, e.g., in Ref.~\cite{Hernandez2018} is not yet possible with our present tools,
%here we probe the nuclear dependency of our results by 
we simply repeat the calculations using a different nuclear Hamiltonian
with two- and three-body interactions derived from chiral effective field theory. 
Following Refs.~\cite{Ji:2018,NND16,Ji13}, we use the two- and three-body interactions
derived in Refs.~\cite{Entem03} and~\cite{Navratil07b}, respectively,  and denote 
results from this Hamiltonian with ``$\chi$EFT''.  Results are shown in Table~\ref{Table_new} 
for different potentials and also for different parameterization of the  nucleonic 
form factor.
 If we use the same nucleon form factor as calculations in previous sections, namely the 
 Kelly form factors, we see that 
the $\chi$EFT interactions shift the electric moments: both $R_E$ and $\ev{R^3_E}_{(2)}$
increase and agree better with the experimental values.
 The dynamical model 
dependency amounts to 1$\%$-2$\%$, which is compatible with the chiral truncation
uncertainty estimate discussed before and is much larger than the sub-percentage few-body or procedure uncertainty. 

% CUT
%\begin{figure}[htb]
%\centering
%\includegraphics[width=0.5\textwidth]{Figure_3He_Ch_Models_bottom_v1.pdf}
%\caption{(Color online) Correlation between the electric moments  of $^3$He: various IA 
%calculations  in comparison with experimental data from Ref.~\cite{Sick2014}. Two different 
%nuclear potentials are used, together with three different parameterizations of the nucleon 
%form factors.  See text for details. }
%\label{Fig_Models}
%\end{figure}

The second variable input we address here is the specific parameterization
of the nucleonic form factors.
Our benchmark calculations are based on the Kelly parameterization, which is widely used due to 
its simplicity and high-quality fit of the available nucleon electromagnetic data. The Kelly 
parameterization yields a proton radius $r_p(K)=0.863(4)$ fm. Another common parameterization
from global fits of electron scattering data is from H\"{o}hler~\cite{HOHLER1976505}. When tested in calculations
of the electric moments of $A=3$~\cite{Piarulli12}, these parameterizations produce results 
in agreement at the sub-percent level. 

Currently, the main uncertainty in this input pertains to the size 
of the proton, stemming from the discrepancy between the determination from muonic hydrogen by 
the CREMA collaboration~\cite{Antognini13}, i.e., $r_p(\mu^-)=0.84087(39)\ {\rm fm}$, 
and the most recent CODATA determination~\cite{CODATA_2014} of
$r_p(e^-)=0.8751(61)\ {\rm fm}$, which does not incorporate the muonic hydrogen result.
In order to conservatively estimate the impact of this discrepancy at the nucleonic level onto
nuclear observables, and in lack of parameterizations that account for this 
proton's size uncertainty in the global fits, we adopt a simple 
parameterization. We the use dipole form to represent the nucleon form factor
as was done, e.g., in Ref.~\cite{FriarPayne:2005_PRC-1_NS_HFS},
fitting the single  parameter
to reproduce either the CREMA or the CODATA proton radius.
Clearly this approximation is completely driven by one observable 
at q=0, whereas the moments are, as we saw, sensitive to the slopes 
and shapes of the nucleonic form factors. 
With this warning in mind, we proceed our analysis. Following Ref.~\cite{Ji:2018}, we take
the neutron electric form factor to be of the modified Galster shape used in~\cite{FriarPayne:2005_PRC-1_NS_HFS}, 
updated to reproduce  $\ev{r^2_E}_n=-0.116$ fm$^2$ from~\cite{PDG_Beringer_2012_PRD}.

Results for the charge radius and third Zemach moment are shown in Table~\ref{Table_new}, where we
observe a 1$\%$--2$\%$ variance, which is as large as the dependency on the nuclear interaction. 
%The results, shown in Figure~\ref{Fig_Models}, demonstrate a large variance that stems from the
%ambiguity at the core of the proton radius puzzle, which is as large as the dependency on 
%the nuclear interaction.
While the specific choice used of nuclear potential and nucleon form factors may significantly 
affect the perceived agreement of the IA calculation with experiment---e.g., the 
 $\chi$EFT potential in combination with the Kelly or CODATA form factor 
is very close to the mean experimental value---we stress that RC and two-body currents 
are missing here. If one added consistently all the uncertainties stemming from the truncation 
in the chiral expansion, all these theoretical points would be statistically in agreement
among themselves and with the experimental values--{\it albeit} with a slightly larger 
but comparable uncertainty. 
%Similar effects are expected in magnetic observables.

\begin{center}
 
\begin{table}
 \caption{Uncertainty budget for the  $^3$He electric moments.
 Experimental values 
are from Ref.~\cite{Sick2014,PURCELL20101}. See text for details. }
 \begin{tabular}{l|c|c|c|c}
                      & $\delta$(Method)     & $\delta$(Dynamics)  & $\delta$(FF) \\
 \hline 
 $R_{E}$              &  0.1\%                    & 0.9\%                          &    0.8\%    \\
 $\ev{R^3_{E}}_{(2)}$ &  0.4-0.7\%                     & 2\%                            &    2.4\%          \\
% $\ev{R^4_{E}}$       &  0.6\% (4\% with $q$-space)&                                &              \\
% $\bra R_{\rm Z}\ket $&                       &                        &                      &              \\
% $R_{M}$              &                       &                        &                      &              \\
% $\ev{\mu}$           &                       &                        &                      &              \\
 \end{tabular}
\label{tab:budget}
\end{table}

\end{center}

Our findings are  summarized in Table~\ref{tab:budget} where we show the uncertainty budget
for these calculations. Here, $\delta$(Method) is the uncertainty from the few-body
method and numerical procedure added in quadrature, $\delta$(Dynamics) is the model dependence
accounted by testing two nuclear Hamiltonians, 
%or changing chiral operator
$\delta$(FF) is the sensitivity
of our results to the use of this single nucleon input. 

As already pointed out,  $\delta$(Method) is
small for $R_E$ and $\ev{R^3_{E}}_{(2)}$ and  
%$\delta$(Dynamics) is estimated in this paper in two ways: 
%once by taking the electromagnetic 
%operator and different orders and evaluating the chiral truncation error, keeping the Hamiltonian fixed,  and once by changing 
%the Hamiltonian while keeping the electromagnetic operator at leading oder. We find that both estimate give roughly a one per cent
%uncertainty and we do not sum them up to avoid possible double-counting.\textcolor{blue}{can we comment on Epelbaum's finding?}
$\delta$(Dynamics) is of the same order as the chiral convergence uncertainty obtained by using the algorithm by Epelbaum {\it et al.}~\cite{Epelbaum2015}.
Finally,  $\delta$(FF) is  roughly estimated using dipole form factors fixed
to reproduce either the CREMA or the CODATA proton radii, giving a very 
conservative uncertainty, also the order of 1$\%$--2$\%$.
It is to be noted that, e.g., the electric charge radii vary by only   
0.15$\%$ when replacing  Kelly nucleon form factors with a different global fit from H\"{o}hler 
{\it et al.}~\cite{HOHLER1976505}, as was done in Ref.~\cite{Piarulli12}

Overall, we observe that  the uncertainty pertaining the
nuclear dynamics and dipole nucleonic form factors are dominant over 
the method uncertainties. 

%However,  method uncertainties
%are observed to  increase for higher order moments, such as $\langle R_E^4 \rangle$, 
%due to the more involved numerical procedures used to evaluate them, as discussed in Section~\ref{sec:Numerical}.

%%%%%%%%%%%%%%%%%%%%%%%%%%%%%%%%%%%%%%%%%%%%%%%%%
\section {Conclusions}
\label{sec:Conclusion}

In this paper, we performed  benchmark calculations 
of electromagnetic moments relevant to ongoing experimental 
efforts, particularly those investigating the spectroscopy
of muon-nucleus systems.

Benchmark calculations in IA are important to assess the reliability  of 
the calculated electromagnetic moments within modern ab initio methods. 
We show that different few-body computational methods lead to compatible results, given the same dynamical inputs. 
We also investigated three distinct numerical procedures
($q$-space, $r$-space, and $qr$-space) that can be used 
to calculate these observables, and have shown that they 
yield comparable results in agreement at the 1$\%$ level or better, 
a part for the fourth electric moment, for which the $q$-space method produces a larger uncertainty.

The dominant source of uncertainty in the calculations is due to the 
employed dynamical inputs, that is, the nuclear Hamiltonian, the
electromagnetic current operators, and the single nucleon parameterizations.  
In particular, few-body and numerical procedure errors are found to 
be at the sub-percent level in calculations of the $^3$He 
electric moments in IA based on the $r$-space procedure.  
The same observables have $\sim 1\%$--$2\%$ variation
when different nuclear Hamiltonians are used. 

We studied the RC and two-body current contributions in the $A=3$
systems using wave functions from the AV18+UIX Hamiltonian, and found that
these contributions are important to reach agreement with the data.
In particular, RC corrections are found to be relevant in electric
moments, while two-body currents are necessary to explain 
magnetic data. The combined contribution from RC and two-body 
currents is at the $3\%$--$5\%$ level in $R_E$ and $\ev{R^3_E}_{(2)}$,
and of the order of $\sim 3\%$--$6\%$ ($\sim 12\%$--$15\%$) 
in $R_Z$ and $R_M$ (the $A=3$ $\mu$'s).
Lastly, in order to make contact with the CREMA findings on
the proton's size, and in order to asses the possible impact of these
findings on nuclear observables, we used a dipole representation
of the nucleonic form factors fitted to reproduce either
the CREMA or the CODATA value. This produces yields a 
rather ample allowance for the uncertainty in the nucleonic input,
and leads to a conservative few-percent error bar on the 
nuclear observables.

This first theoretical study of electromagnetic moments  indicates that 
the total theoretical uncertainty is  of the same order of magnitude 
as the experimental one at least for the charge, magnetic and Zemach 
radii, third Zemach moment, and magnetic moments.
Finally, we remark that, while currently theoretical uncertainties 
seem comparable to those of electron scattering  data,
the anticipated precision of muonic experiments will be superior, 
further challenging the theory.
Performing a fully consistent calculation accompanied by a thorough 
statistical and systematical analysis of these observables
is demanding and will be explored in our future studies. 

\vspace{0.5cm}
{\it Acknowledgments --}  S.P. and M.P. would like to thank Rocco 
Schiavilla and Laura Elisa Marcucci
for useful discussions, and gratefully acknowledge the computing resources 
of the high-performance computing cluster operated 
by the Laboratory Computing Resource Center (LCRC) at Argonne National 
Laboratory (ANL). 
This work was supported in parts
by the Natural Sciences and Engineering Research Council (NSERC), the
National Research Council of Canada, by the Deutsche
Forschungsgemeinschaft DFG through the Collaborative Research Center
[The Low-Energy Frontier of the Standard Model (SFB 1044)], and
through the Cluster of Excellence [Precision Physics, Fundamental
  Interactions and Structure of Matter (PRISMA)]. The work of R.B.W. is
supported by the U.S. Department of Energy, Office of Science, Office of Nuclear
Physics, under Contract No. DE-AC02-06CH11357.

\bibliography{bibliography_zemach}

%merlin.mbs apsrev4-1.bst 2010-07-25 4.21a (PWD, AO, DPC) hacked
%Control: key (0)
%Control: author (8) initials jnrlst
%Control: editor formatted (1) identically to author
%Control: production of article title (-1) disabled
%Control: page (0) single
%Control: year (1) truncated
%Control: production of eprint (0) enabled
\begin{thebibliography}{67}%
\makeatletter
\providecommand \@ifxundefined [1]{%
 \@ifx{#1\undefined}
}%
\providecommand \@ifnum [1]{%
 \ifnum #1\expandafter \@firstoftwo
 \else \expandafter \@secondoftwo
 \fi
}%
\providecommand \@ifx [1]{%
 \ifx #1\expandafter \@firstoftwo
 \else \expandafter \@secondoftwo
 \fi
}%
\providecommand \natexlab [1]{#1}%
\providecommand \enquote  [1]{``#1''}%
\providecommand \bibnamefont  [1]{#1}%
\providecommand \bibfnamefont [1]{#1}%
\providecommand \citenamefont [1]{#1}%
\providecommand \href@noop [0]{\@secondoftwo}%
\providecommand \href [0]{\begingroup \@sanitize@url \@href}%
\providecommand \@href[1]{\@@startlink{#1}\@@href}%
\providecommand \@@href[1]{\endgroup#1\@@endlink}%
\providecommand \@sanitize@url [0]{\catcode `\\12\catcode `\$12\catcode
  `\&12\catcode `\#12\catcode `\^12\catcode `\_12\catcode `\%12\relax}%
\providecommand \@@startlink[1]{}%
\providecommand \@@endlink[0]{}%
\providecommand \url  [0]{\begingroup\@sanitize@url \@url }%
\providecommand \@url [1]{\endgroup\@href {#1}{\urlprefix }}%
\providecommand \urlprefix  [0]{URL }%
\providecommand \Eprint [0]{\href }%
\providecommand \doibase [0]{http://dx.doi.org/}%
\providecommand \selectlanguage [0]{\@gobble}%
\providecommand \bibinfo  [0]{\@secondoftwo}%
\providecommand \bibfield  [0]{\@secondoftwo}%
\providecommand \translation [1]{[#1]}%
\providecommand \BibitemOpen [0]{}%
\providecommand \bibitemStop [0]{}%
\providecommand \bibitemNoStop [0]{.\EOS\space}%
\providecommand \EOS [0]{\spacefactor3000\relax}%
\providecommand \BibitemShut  [1]{\csname bibitem#1\endcsname}%
\let\auto@bib@innerbib\@empty
%</preamble>
\bibitem [{\citenamefont {Pohl}\ \emph {et~al.}(2010)\citenamefont {Pohl} \emph
  {et~al.}}]{Pohl:2010zza}%
  \BibitemOpen
  \bibfield  {author} {\bibinfo {author} {\bibfnamefont {R.}~\bibnamefont
  {Pohl}} \emph {et~al.},\ }\href {\doibase 10.1038/nature09250} {\bibfield
  {journal} {\bibinfo  {journal} {Nature}\ }\textbf {\bibinfo {volume} {466}},\
  \bibinfo {pages} {213} (\bibinfo {year} {2010})}\BibitemShut {NoStop}%
\bibitem [{\citenamefont {Antognini}\ \emph {et~al.}(2013)\citenamefont
  {Antognini} \emph {et~al.}}]{Antognini13}%
  \BibitemOpen
  \bibfield  {author} {\bibinfo {author} {\bibfnamefont {A.}~\bibnamefont
  {Antognini}} \emph {et~al.},\ }\href {\doibase 10.1126/science.1230016}
  {\bibfield  {journal} {\bibinfo  {journal} {Science}\ }\textbf {\bibinfo
  {volume} {339}},\ \bibinfo {pages} {417} (\bibinfo {year}
  {2013})}\BibitemShut {NoStop}%
\bibitem [{\citenamefont {Pohl}\ \emph {et~al.}(2016)\citenamefont {Pohl},
  \citenamefont {Nez}, \citenamefont {Fernandes}, \citenamefont {Amaro},
  \citenamefont {Biraben}, \citenamefont {Cardoso}, \citenamefont {Covita},
  \citenamefont {Dax}, \citenamefont {Dhawan}, \citenamefont {Diepold},
  \citenamefont {Giesen}, \citenamefont {Gouvea}, \citenamefont {Graf},
  \citenamefont {H{\"a}nsch}, \citenamefont {Indelicato}, \citenamefont
  {Julien}, \citenamefont {Knowles}, \citenamefont {Kottmann}, \citenamefont
  {Le~Bigot}, \citenamefont {Liu}, \citenamefont {Lopes}, \citenamefont
  {Ludhova}, \citenamefont {Monteiro}, \citenamefont {Mulhauser}, \citenamefont
  {Nebel}, \citenamefont {Rabinowitz}, \citenamefont {dos Santos},
  \citenamefont {Schaller}, \citenamefont {Schuhmann}, \citenamefont {Schwob},
  \citenamefont {Taqqu}, \citenamefont {Veloso}, \citenamefont {Antognini},\
  and\ }]{Pohl_2016_Science}%
  \BibitemOpen
  \bibfield  {author} {\bibinfo {author} {\bibfnamefont {R.}~\bibnamefont
  {Pohl}}, \bibinfo {author} {\bibfnamefont {F.}~\bibnamefont {Nez}}, \bibinfo
  {author} {\bibfnamefont {L.~M.~P.}\ \bibnamefont {Fernandes}}, \bibinfo
  {author} {\bibfnamefont {F.~D.}\ \bibnamefont {Amaro}}, \bibinfo {author}
  {\bibfnamefont {F.}~\bibnamefont {Biraben}}, \bibinfo {author} {\bibfnamefont
  {J.~M.~R.}\ \bibnamefont {Cardoso}}, \bibinfo {author} {\bibfnamefont
  {D.~S.}\ \bibnamefont {Covita}}, \bibinfo {author} {\bibfnamefont
  {A.}~\bibnamefont {Dax}}, \bibinfo {author} {\bibfnamefont {S.}~\bibnamefont
  {Dhawan}}, \bibinfo {author} {\bibfnamefont {M.}~\bibnamefont {Diepold}},
  \bibinfo {author} {\bibfnamefont {A.}~\bibnamefont {Giesen}}, \bibinfo
  {author} {\bibfnamefont {A.~L.}\ \bibnamefont {Gouvea}}, \bibinfo {author}
  {\bibfnamefont {T.}~\bibnamefont {Graf}}, \bibinfo {author} {\bibfnamefont
  {T.~W.}\ \bibnamefont {H{\"a}nsch}}, \bibinfo {author} {\bibfnamefont
  {P.}~\bibnamefont {Indelicato}}, \bibinfo {author} {\bibfnamefont
  {L.}~\bibnamefont {Julien}}, \bibinfo {author} {\bibfnamefont
  {P.}~\bibnamefont {Knowles}}, \bibinfo {author} {\bibfnamefont
  {F.}~\bibnamefont {Kottmann}}, \bibinfo {author} {\bibfnamefont {E.-O.}\
  \bibnamefont {Le~Bigot}}, \bibinfo {author} {\bibfnamefont {Y.-W.}\
  \bibnamefont {Liu}}, \bibinfo {author} {\bibfnamefont {J.~A.~M.}\
  \bibnamefont {Lopes}}, \bibinfo {author} {\bibfnamefont {L.}~\bibnamefont
  {Ludhova}}, \bibinfo {author} {\bibfnamefont {C.~M.~B.}\ \bibnamefont
  {Monteiro}}, \bibinfo {author} {\bibfnamefont {F.}~\bibnamefont {Mulhauser}},
  \bibinfo {author} {\bibfnamefont {T.}~\bibnamefont {Nebel}}, \bibinfo
  {author} {\bibfnamefont {P.}~\bibnamefont {Rabinowitz}}, \bibinfo {author}
  {\bibfnamefont {J.~M.~F.}\ \bibnamefont {dos Santos}}, \bibinfo {author}
  {\bibfnamefont {L.~A.}\ \bibnamefont {Schaller}}, \bibinfo {author}
  {\bibfnamefont {K.}~\bibnamefont {Schuhmann}}, \bibinfo {author}
  {\bibfnamefont {C.}~\bibnamefont {Schwob}}, \bibinfo {author} {\bibfnamefont
  {D.}~\bibnamefont {Taqqu}}, \bibinfo {author} {\bibfnamefont {J.~F. C.~A.}\
  \bibnamefont {Veloso}}, \bibinfo {author} {\bibfnamefont {A.}~\bibnamefont
  {Antognini}}, \ and\ ,\ }\href {\doibase 10.1126/science.aaf2468} {\bibfield
  {journal} {\bibinfo  {journal} {Science}\ }\textbf {\bibinfo {volume}
  {353}},\ \bibinfo {pages} {669} (\bibinfo {year} {2016})},\ \Eprint
  {http://arxiv.org/abs/http://science.sciencemag.org/content/353/6300/669.full.pdf}
  {http://science.sciencemag.org/content/353/6300/669.full.pdf} \BibitemShut
  {NoStop}%
\bibitem [{\citenamefont {Pohl}\ \emph {et~al.}(2017)\citenamefont {Pohl},
  \citenamefont {Nez}, \citenamefont {Udem}, \citenamefont {Antognini},
  \citenamefont {Beyer}, \citenamefont {Fleurbaey}, \citenamefont {Grinin},
  \citenamefont {H{\"a}nsch}, \citenamefont {Julien}, \citenamefont {Kottmann},
  \citenamefont {Krauth}, \citenamefont {Maisenbacher}, \citenamefont
  {Matveev},\ and\ \citenamefont {Biraben}}]{Metrologia2017}%
  \BibitemOpen
  \bibfield  {author} {\bibinfo {author} {\bibfnamefont {R.}~\bibnamefont
  {Pohl}}, \bibinfo {author} {\bibfnamefont {F.}~\bibnamefont {Nez}}, \bibinfo
  {author} {\bibfnamefont {T.}~\bibnamefont {Udem}}, \bibinfo {author}
  {\bibfnamefont {A.}~\bibnamefont {Antognini}}, \bibinfo {author}
  {\bibfnamefont {A.}~\bibnamefont {Beyer}}, \bibinfo {author} {\bibfnamefont
  {H.}~\bibnamefont {Fleurbaey}}, \bibinfo {author} {\bibfnamefont
  {A.}~\bibnamefont {Grinin}}, \bibinfo {author} {\bibfnamefont {T.~W.}\
  \bibnamefont {H{\"a}nsch}}, \bibinfo {author} {\bibfnamefont
  {L.}~\bibnamefont {Julien}}, \bibinfo {author} {\bibfnamefont
  {F.}~\bibnamefont {Kottmann}}, \bibinfo {author} {\bibfnamefont {J.~J.}\
  \bibnamefont {Krauth}}, \bibinfo {author} {\bibfnamefont {L.}~\bibnamefont
  {Maisenbacher}}, \bibinfo {author} {\bibfnamefont {A.}~\bibnamefont
  {Matveev}}, \ and\ \bibinfo {author} {\bibfnamefont {F.}~\bibnamefont
  {Biraben}},\ }\href {http://stacks.iop.org/0026-1394/54/i=2/a=L1} {\bibfield
  {journal} {\bibinfo  {journal} {Metrologia}\ }\textbf {\bibinfo {volume}
  {54}},\ \bibinfo {pages} {L1} (\bibinfo {year} {2017})}\BibitemShut {NoStop}%
\bibitem [{\citenamefont {Bernauer}\ and\ \citenamefont
  {Pohl}(2014)}]{Bernauer_Pohl}%
  \BibitemOpen
  \bibfield  {author} {\bibinfo {author} {\bibfnamefont {J.~C.}\ \bibnamefont
  {Bernauer}}\ and\ \bibinfo {author} {\bibfnamefont {R.}~\bibnamefont
  {Pohl}},\ }\href {\doibase 10.1038/scientificamerican0214-32} {\bibfield
  {journal} {\bibinfo  {journal} {Sci. Am.}\ }\textbf {\bibinfo {volume}
  {310}},\ \bibinfo {pages} {18} (\bibinfo {year} {2014})}\BibitemShut
  {NoStop}%
%%CITATION = SCAMA,310,18;%%
\bibitem [{\citenamefont {Pohl}\ \emph {et~al.}(2013)\citenamefont {Pohl},
  \citenamefont {Gilman}, \citenamefont {Miller},\ and\ \citenamefont
  {Pachucki}}]{Pohl_Review}%
  \BibitemOpen
  \bibfield  {author} {\bibinfo {author} {\bibfnamefont {R.}~\bibnamefont
  {Pohl}}, \bibinfo {author} {\bibfnamefont {R.}~\bibnamefont {Gilman}},
  \bibinfo {author} {\bibfnamefont {G.~A.}\ \bibnamefont {Miller}}, \ and\
  \bibinfo {author} {\bibfnamefont {K.}~\bibnamefont {Pachucki}},\ }\href
  {\doibase 10.1146/annurev-nucl-102212-170627} {\bibfield  {journal} {\bibinfo
   {journal} {Ann. Rev. Nucl. Part. Sci.}\ }\textbf {\bibinfo {volume} {63}},\
  \bibinfo {pages} {175} (\bibinfo {year} {2013})},\ \Eprint
  {http://arxiv.org/abs/1301.0905} {arXiv:1301.0905 [physics.atom-ph]}
  \BibitemShut {NoStop}%
%%CITATION = ARXIV:1301.0905;%%
\bibitem [{\citenamefont {Beyer}\ \emph {et~al.}(2017)\citenamefont {Beyer}
  \emph {et~al.}}]{Beyer79}%
  \BibitemOpen
  \bibfield  {author} {\bibinfo {author} {\bibfnamefont {A.}~\bibnamefont
  {Beyer}} \emph {et~al.},\ }\href {\doibase 10.1126/science.aah6677}
  {\bibfield  {journal} {\bibinfo  {journal} {Science}\ }\textbf {\bibinfo
  {volume} {358}},\ \bibinfo {pages} {79} (\bibinfo {year} {2017})}\BibitemShut
  {NoStop}%
\bibitem [{\citenamefont {Fleurbaey}\ \emph {et~al.}(2018)\citenamefont
  {Fleurbaey} \emph {et~al.}}]{Paris}%
  \BibitemOpen
  \bibfield  {author} {\bibinfo {author} {\bibfnamefont {H.}~\bibnamefont
  {Fleurbaey}} \emph {et~al.},\ }\href {\doibase
  10.1103/PhysRevLett.120.183001} {\bibfield  {journal} {\bibinfo  {journal}
  {Phys. Rev. Lett.}\ }\textbf {\bibinfo {volume} {120}},\ \bibinfo {pages}
  {183001} (\bibinfo {year} {2018})}\BibitemShut {NoStop}%
\bibitem [{\citenamefont {Pohl}\ \emph {et~al.}()\citenamefont {Pohl},
  \citenamefont {Nez}, \citenamefont {Fernandes}, \citenamefont {Ahmed},
  \citenamefont {Amaro}, \citenamefont {Amaro}, \citenamefont {Biraben},
  \citenamefont {Cardoso}, \citenamefont {Covita}, \citenamefont {Dax},
  \citenamefont {Dhawan}, \citenamefont {Diepold}, \citenamefont {Franke},
  \citenamefont {Galtier}, \citenamefont {Giesen}, \citenamefont {Gouvea},
  \citenamefont {G{\"o}tzfried}, \citenamefont {Graf}, \citenamefont
  {H{\"a}nsch}, \citenamefont {Hildebrandt}, \citenamefont {Indelicato},
  \citenamefont {Julien}, \citenamefont {Kirch}, \citenamefont {Knecht},
  \citenamefont {Knowles}, \citenamefont {Kottmann}, \citenamefont {Krauth},
  \citenamefont {Bigot}, \citenamefont {Liu}, \citenamefont {Lopes},
  \citenamefont {Ludhova}, \citenamefont {Machado}, \citenamefont {Monteiro},
  \citenamefont {Mulhauser}, \citenamefont {Nebel}, \citenamefont {Rabinowitz},
  \citenamefont {dos Santos}, \citenamefont {Santos}, \citenamefont {Schaller},
  \citenamefont {Schuhmann}, \citenamefont {Schwob}, \citenamefont {Szabo},
  \citenamefont {Taqqu}, \citenamefont {Veloso}, \citenamefont {Voss},
  \citenamefont {Weichelt},\ and\ \citenamefont {Antognini}}]{Pohl_2016_LEAP}%
  \BibitemOpen
  \bibfield  {author} {\bibinfo {author} {\bibfnamefont {R.}~\bibnamefont
  {Pohl}}, \bibinfo {author} {\bibfnamefont {F.}~\bibnamefont {Nez}}, \bibinfo
  {author} {\bibfnamefont {L.~M.~P.}\ \bibnamefont {Fernandes}}, \bibinfo
  {author} {\bibfnamefont {M.~A.}\ \bibnamefont {Ahmed}}, \bibinfo {author}
  {\bibfnamefont {F.~D.}\ \bibnamefont {Amaro}}, \bibinfo {author}
  {\bibfnamefont {P.}~\bibnamefont {Amaro}}, \bibinfo {author} {\bibfnamefont
  {F.}~\bibnamefont {Biraben}}, \bibinfo {author} {\bibfnamefont {J.~M.~R.}\
  \bibnamefont {Cardoso}}, \bibinfo {author} {\bibfnamefont {D.~S.}\
  \bibnamefont {Covita}}, \bibinfo {author} {\bibfnamefont {A.}~\bibnamefont
  {Dax}}, \bibinfo {author} {\bibfnamefont {S.}~\bibnamefont {Dhawan}},
  \bibinfo {author} {\bibfnamefont {M.}~\bibnamefont {Diepold}}, \bibinfo
  {author} {\bibfnamefont {B.}~\bibnamefont {Franke}}, \bibinfo {author}
  {\bibfnamefont {S.}~\bibnamefont {Galtier}}, \bibinfo {author} {\bibfnamefont
  {A.}~\bibnamefont {Giesen}}, \bibinfo {author} {\bibfnamefont {A.~L.}\
  \bibnamefont {Gouvea}}, \bibinfo {author} {\bibfnamefont {J.}~\bibnamefont
  {G{\"o}tzfried}}, \bibinfo {author} {\bibfnamefont {T.}~\bibnamefont {Graf}},
  \bibinfo {author} {\bibfnamefont {T.~W.}\ \bibnamefont {H{\"a}nsch}},
  \bibinfo {author} {\bibfnamefont {M.}~\bibnamefont {Hildebrandt}}, \bibinfo
  {author} {\bibfnamefont {P.}~\bibnamefont {Indelicato}}, \bibinfo {author}
  {\bibfnamefont {L.}~\bibnamefont {Julien}}, \bibinfo {author} {\bibfnamefont
  {K.}~\bibnamefont {Kirch}}, \bibinfo {author} {\bibfnamefont
  {A.}~\bibnamefont {Knecht}}, \bibinfo {author} {\bibfnamefont
  {P.}~\bibnamefont {Knowles}}, \bibinfo {author} {\bibfnamefont
  {F.}~\bibnamefont {Kottmann}}, \bibinfo {author} {\bibfnamefont {J.~J.}\
  \bibnamefont {Krauth}}, \bibinfo {author} {\bibfnamefont {E.-O.~L.}\
  \bibnamefont {Bigot}}, \bibinfo {author} {\bibfnamefont {Y.-W.}\ \bibnamefont
  {Liu}}, \bibinfo {author} {\bibfnamefont {J.~A.~M.}\ \bibnamefont {Lopes}},
  \bibinfo {author} {\bibfnamefont {L.}~\bibnamefont {Ludhova}}, \bibinfo
  {author} {\bibfnamefont {J.}~\bibnamefont {Machado}}, \bibinfo {author}
  {\bibfnamefont {C.~M.~B.}\ \bibnamefont {Monteiro}}, \bibinfo {author}
  {\bibfnamefont {F.}~\bibnamefont {Mulhauser}}, \bibinfo {author}
  {\bibfnamefont {T.}~\bibnamefont {Nebel}}, \bibinfo {author} {\bibfnamefont
  {P.}~\bibnamefont {Rabinowitz}}, \bibinfo {author} {\bibfnamefont {J.~M.~F.}\
  \bibnamefont {dos Santos}}, \bibinfo {author} {\bibfnamefont {J.~P.}\
  \bibnamefont {Santos}}, \bibinfo {author} {\bibfnamefont {L.~A.}\
  \bibnamefont {Schaller}}, \bibinfo {author} {\bibfnamefont {K.}~\bibnamefont
  {Schuhmann}}, \bibinfo {author} {\bibfnamefont {C.}~\bibnamefont {Schwob}},
  \bibinfo {author} {\bibfnamefont {C.~I.}\ \bibnamefont {Szabo}}, \bibinfo
  {author} {\bibfnamefont {D.}~\bibnamefont {Taqqu}}, \bibinfo {author}
  {\bibfnamefont {J.~F. C.~A.}\ \bibnamefont {Veloso}}, \bibinfo {author}
  {\bibfnamefont {A.}~\bibnamefont {Voss}}, \bibinfo {author} {\bibfnamefont
  {B.}~\bibnamefont {Weichelt}}, \ and\ \bibinfo {author} {\bibfnamefont
  {A.}~\bibnamefont {Antognini}},\ }\enquote {\bibinfo {title} {Laser
  spectroscopy of muonic atoms and ions},}\ in\ \href {\doibase
  10.7566/JPSCP.18.011021} {\emph {\bibinfo {booktitle} {Proceedings of the
  12th International Conference on Low Energy Antiproton Physics
  (LEAP2016)}}},\ \Eprint
  {http://arxiv.org/abs/https://journals.jps.jp/doi/pdf/10.7566/JPSCP.18.011021}
  {https://journals.jps.jp/doi/pdf/10.7566/JPSCP.18.011021} \BibitemShut
  {NoStop}%
\bibitem [{\citenamefont {Ji}\ \emph {et~al.}(2018)\citenamefont {Ji},
  \citenamefont {Bacca}, \citenamefont {Barnea}, \citenamefont {Hernandez},\
  and\ \citenamefont {{Nevo Dinur}}}]{Ji:2018}%
  \BibitemOpen
  \bibfield  {author} {\bibinfo {author} {\bibfnamefont {C.}~\bibnamefont
  {Ji}}, \bibinfo {author} {\bibfnamefont {S.}~\bibnamefont {Bacca}}, \bibinfo
  {author} {\bibfnamefont {N.}~\bibnamefont {Barnea}}, \bibinfo {author}
  {\bibfnamefont {O.~J.}\ \bibnamefont {Hernandez}}, \ and\ \bibinfo {author}
  {\bibfnamefont {N.}~\bibnamefont {{Nevo Dinur}}},\ }\href
  {http://stacks.iop.org/0954-3899/45/i=9/a=093002} {\bibfield  {journal}
  {\bibinfo  {journal} {Journal of Physics G: Nuclear and Particle Physics}\
  }\textbf {\bibinfo {volume} {45}},\ \bibinfo {pages} {093002} (\bibinfo
  {year} {2018})}\BibitemShut {NoStop}%
\bibitem [{\citenamefont {Franke}\ \emph {et~al.}(2017)\citenamefont {Franke},
  \citenamefont {Krauth}, \citenamefont {Antognini}, \citenamefont {Diepold},
  \citenamefont {Kottmann},\ and\ \citenamefont {Pohl}}]{Franke_2017}%
  \BibitemOpen
  \bibfield  {author} {\bibinfo {author} {\bibfnamefont {B.}~\bibnamefont
  {Franke}}, \bibinfo {author} {\bibfnamefont {J.~J.}\ \bibnamefont {Krauth}},
  \bibinfo {author} {\bibfnamefont {A.}~\bibnamefont {Antognini}}, \bibinfo
  {author} {\bibfnamefont {M.}~\bibnamefont {Diepold}}, \bibinfo {author}
  {\bibfnamefont {F.}~\bibnamefont {Kottmann}}, \ and\ \bibinfo {author}
  {\bibfnamefont {R.}~\bibnamefont {Pohl}},\ }\href {\doibase
  10.1140/epjd/e2017-80296-1} {\bibfield  {journal} {\bibinfo  {journal} {The
  European Physical Journal D}\ }\textbf {\bibinfo {volume} {71}},\ \bibinfo
  {pages} {341} (\bibinfo {year} {2017})}\BibitemShut {NoStop}%
\bibitem [{\citenamefont {Diepold}\ \emph {et~al.}(2018)\citenamefont
  {Diepold}, \citenamefont {Franke}, \citenamefont {Krauth}, \citenamefont
  {Antognini}, \citenamefont {Kottmann},\ and\ \citenamefont
  {Pohl}}]{Diepold_2018}%
  \BibitemOpen
  \bibfield  {author} {\bibinfo {author} {\bibfnamefont {M.}~\bibnamefont
  {Diepold}}, \bibinfo {author} {\bibfnamefont {B.}~\bibnamefont {Franke}},
  \bibinfo {author} {\bibfnamefont {J.~J.}\ \bibnamefont {Krauth}}, \bibinfo
  {author} {\bibfnamefont {A.}~\bibnamefont {Antognini}}, \bibinfo {author}
  {\bibfnamefont {F.}~\bibnamefont {Kottmann}}, \ and\ \bibinfo {author}
  {\bibfnamefont {R.}~\bibnamefont {Pohl}},\ }\href {\doibase
  https://doi.org/10.1016/j.aop.2018.07.015} {\bibfield  {journal} {\bibinfo
  {journal} {Annals of Physics}\ }\textbf {\bibinfo {volume} {396}},\ \bibinfo
  {pages} {220 } (\bibinfo {year} {2018})}\BibitemShut {NoStop}%
\bibitem [{\citenamefont {Eides}\ \emph {et~al.}(2001)\citenamefont {Eides},
  \citenamefont {Grotch},\ and\ \citenamefont {Shelyuto}}]{Eides_PR2001}%
  \BibitemOpen
  \bibfield  {author} {\bibinfo {author} {\bibfnamefont {M.~I.}\ \bibnamefont
  {Eides}}, \bibinfo {author} {\bibfnamefont {H.}~\bibnamefont {Grotch}}, \
  and\ \bibinfo {author} {\bibfnamefont {V.~A.}\ \bibnamefont {Shelyuto}},\
  }\href {\doibase https://doi.org/10.1016/S0370-1573(00)00077-6} {\bibfield
  {journal} {\bibinfo  {journal} {Physics Reports}\ }\textbf {\bibinfo {volume}
  {342}},\ \bibinfo {pages} {63 } (\bibinfo {year} {2001})}\BibitemShut
  {NoStop}%
\bibitem [{\citenamefont {Borie}(2012)}]{Borie2012_AoP_arXiv}%
  \BibitemOpen
  \bibfield  {author} {\bibinfo {author} {\bibfnamefont {E.}~\bibnamefont
  {Borie}},\ }\href {\doibase http://dx.doi.org/10.1016/j.aop.2011.11.017}
  {\bibfield  {journal} {\bibinfo  {journal} {Annals of Physics}\ }\textbf
  {\bibinfo {volume} {327}},\ \bibinfo {pages} {733 } (\bibinfo {year}
  {2012})}\BibitemShut {NoStop}%
\bibitem [{\citenamefont {Korzinin}\ \emph {et~al.}(2018)\citenamefont
  {Korzinin}, \citenamefont {Shelyuto}, \citenamefont {Ivanov},\ and\
  \citenamefont {Karshenboim}}]{Korzinin_2018}%
  \BibitemOpen
  \bibfield  {author} {\bibinfo {author} {\bibfnamefont {E.~Y.}\ \bibnamefont
  {Korzinin}}, \bibinfo {author} {\bibfnamefont {V.~A.}\ \bibnamefont
  {Shelyuto}}, \bibinfo {author} {\bibfnamefont {V.~G.}\ \bibnamefont
  {Ivanov}}, \ and\ \bibinfo {author} {\bibfnamefont {S.~G.}\ \bibnamefont
  {Karshenboim}},\ }\href {\doibase 10.1103/PhysRevA.97.012514} {\bibfield
  {journal} {\bibinfo  {journal} {Phys. Rev. A}\ }\textbf {\bibinfo {volume}
  {97}},\ \bibinfo {pages} {012514} (\bibinfo {year} {2018})}\BibitemShut
  {NoStop}%
\bibitem [{\citenamefont {Kalinowski}\ \emph {et~al.}(2018)\citenamefont
  {Kalinowski}, \citenamefont {Pachucki},\ and\ \citenamefont
  {Yerokhin}}]{Marcin}%
  \BibitemOpen
  \bibfield  {author} {\bibinfo {author} {\bibfnamefont {M.}~\bibnamefont
  {Kalinowski}}, \bibinfo {author} {\bibfnamefont {K.}~\bibnamefont
  {Pachucki}}, \ and\ \bibinfo {author} {\bibfnamefont {V.~A.}\ \bibnamefont
  {Yerokhin}},\ }\href@noop {} {\  (\bibinfo {year} {2018})},\ \Eprint
  {http://arxiv.org/abs/1810.06601} {arXiv:1810.06601 [physics.atom-ph]}
  \BibitemShut {NoStop}%
%%CITATION = ARXIV:1810.06601;%%
\bibitem [{\citenamefont {Pachucki}(2011)}]{Pachucki2011}%
  \BibitemOpen
  \bibfield  {author} {\bibinfo {author} {\bibfnamefont {K.}~\bibnamefont
  {Pachucki}},\ }\href {\doibase 10.1103/PhysRevLett.106.193007} {\bibfield
  {journal} {\bibinfo  {journal} {Phys. Rev. Lett.}\ }\textbf {\bibinfo
  {volume} {106}},\ \bibinfo {pages} {193007} (\bibinfo {year}
  {2011})}\BibitemShut {NoStop}%
\bibitem [{\citenamefont {Friar}(2013)}]{Friar:2013rha}%
  \BibitemOpen
  \bibfield  {author} {\bibinfo {author} {\bibfnamefont {J.~L.}\ \bibnamefont
  {Friar}},\ }\href {\doibase 10.1103/PhysRevC.88.034003} {\bibfield  {journal}
  {\bibinfo  {journal} {Phys. Rev. C}\ }\textbf {\bibinfo {volume} {88}},\
  \bibinfo {pages} {034003} (\bibinfo {year} {2013})}\BibitemShut {NoStop}%
\bibitem [{\citenamefont {Ji}\ \emph {et~al.}(2013)\citenamefont {Ji},
  \citenamefont {{Nevo Dinur}}, \citenamefont {Bacca},\ and\ \citenamefont
  {Barnea}}]{Ji13}%
  \BibitemOpen
  \bibfield  {author} {\bibinfo {author} {\bibfnamefont {C.}~\bibnamefont
  {Ji}}, \bibinfo {author} {\bibfnamefont {N.}~\bibnamefont {{Nevo Dinur}}},
  \bibinfo {author} {\bibfnamefont {S.}~\bibnamefont {Bacca}}, \ and\ \bibinfo
  {author} {\bibfnamefont {N.}~\bibnamefont {Barnea}},\ }\href {\doibase
  10.1103/PhysRevLett.111.143402} {\bibfield  {journal} {\bibinfo  {journal}
  {Phys.\ Rev.\ Lett.}\ }\textbf {\bibinfo {volume} {111}},\ \bibinfo {pages}
  {143402} (\bibinfo {year} {2013})}\BibitemShut {NoStop}%
\bibitem [{\citenamefont {Carlson}\ \emph {et~al.}(2014)\citenamefont
  {Carlson}, \citenamefont {Gorchtein},\ and\ \citenamefont
  {Vanderhaeghen}}]{Carlson:2013xea}%
  \BibitemOpen
  \bibfield  {author} {\bibinfo {author} {\bibfnamefont {C.~E.}\ \bibnamefont
  {Carlson}}, \bibinfo {author} {\bibfnamefont {M.}~\bibnamefont {Gorchtein}},
  \ and\ \bibinfo {author} {\bibfnamefont {M.}~\bibnamefont {Vanderhaeghen}},\
  }\href {\doibase 10.1103/PhysRevA.89.022504} {\bibfield  {journal} {\bibinfo
  {journal} {Phys. Rev. A}\ }\textbf {\bibinfo {volume} {89}},\ \bibinfo
  {pages} {022504} (\bibinfo {year} {2014})}\BibitemShut {NoStop}%
\bibitem [{\citenamefont {Hernandez}\ \emph {et~al.}(2014)\citenamefont
  {Hernandez}, \citenamefont {Ji}, \citenamefont {Bacca}, \citenamefont {{Nevo
  Dinur}},\ and\ \citenamefont {Barnea}}]{Hernandez14}%
  \BibitemOpen
  \bibfield  {author} {\bibinfo {author} {\bibfnamefont {O.}~\bibnamefont
  {Hernandez}}, \bibinfo {author} {\bibfnamefont {C.}~\bibnamefont {Ji}},
  \bibinfo {author} {\bibfnamefont {S.}~\bibnamefont {Bacca}}, \bibinfo
  {author} {\bibfnamefont {N.}~\bibnamefont {{Nevo Dinur}}}, \ and\ \bibinfo
  {author} {\bibfnamefont {N.}~\bibnamefont {Barnea}},\ }\href
  {http://www.sciencedirect.com/science/article/pii/S0370269314005413}
  {\bibfield  {journal} {\bibinfo  {journal} {Physics Letters B}\ }\textbf
  {\bibinfo {volume} {736}},\ \bibinfo {pages} {344} (\bibinfo {year}
  {2014})}\BibitemShut {NoStop}%
\bibitem [{\citenamefont {Pachucki}\ and\ \citenamefont
  {Wienczek}(2015)}]{Pachucki2015}%
  \BibitemOpen
  \bibfield  {author} {\bibinfo {author} {\bibfnamefont {K.}~\bibnamefont
  {Pachucki}}\ and\ \bibinfo {author} {\bibfnamefont {A.}~\bibnamefont
  {Wienczek}},\ }\href {\doibase 10.1103/PhysRevA.91.040503} {\bibfield
  {journal} {\bibinfo  {journal} {Phys. Rev.}\ }\textbf {\bibinfo {volume}
  {A91}},\ \bibinfo {pages} {040503} (\bibinfo {year} {2015})},\ \Eprint
  {http://arxiv.org/abs/1501.07451} {arXiv:1501.07451 [physics.atom-ph]}
  \BibitemShut {NoStop}%
%%CITATION = ARXIV:1501.07451;%%
\bibitem [{\citenamefont {{Nevo Dinur}}\ \emph {et~al.}(2016)\citenamefont
  {{Nevo Dinur}}, \citenamefont {Ji}, \citenamefont {Bacca},\ and\
  \citenamefont {Barnea}}]{NND16}%
  \BibitemOpen
  \bibfield  {author} {\bibinfo {author} {\bibfnamefont {N.}~\bibnamefont
  {{Nevo Dinur}}}, \bibinfo {author} {\bibfnamefont {C.}~\bibnamefont {Ji}},
  \bibinfo {author} {\bibfnamefont {S.}~\bibnamefont {Bacca}}, \ and\ \bibinfo
  {author} {\bibfnamefont {N.}~\bibnamefont {Barnea}},\ }\href {\doibase
  http://dx.doi.org/10.1016/j.physletb.2016.02.023} {\bibfield  {journal}
  {\bibinfo  {journal} {Physics Letters B}\ }\textbf {\bibinfo {volume}
  {755}},\ \bibinfo {pages} {380} (\bibinfo {year} {2016})}\BibitemShut
  {NoStop}%
\bibitem [{\citenamefont {Carlson}\ \emph {et~al.}(2017)\citenamefont
  {Carlson}, \citenamefont {Gorchtein},\ and\ \citenamefont
  {Vanderhaeghen}}]{Carlson:2016cii}%
  \BibitemOpen
  \bibfield  {author} {\bibinfo {author} {\bibfnamefont {C.~E.}\ \bibnamefont
  {Carlson}}, \bibinfo {author} {\bibfnamefont {M.}~\bibnamefont {Gorchtein}},
  \ and\ \bibinfo {author} {\bibfnamefont {M.}~\bibnamefont {Vanderhaeghen}},\
  }\href {\doibase 10.1103/PhysRevA.95.012506} {\bibfield  {journal} {\bibinfo
  {journal} {Phys. Rev.}\ }\textbf {\bibinfo {volume} {A95}},\ \bibinfo {pages}
  {012506} (\bibinfo {year} {2017})},\ \Eprint
  {http://arxiv.org/abs/1611.06192} {arXiv:1611.06192 [nucl-th]} \BibitemShut
  {NoStop}%
%%CITATION = ARXIV:1611.06192;%%
\bibitem [{\citenamefont {Kamada}\ \emph {et~al.}(2001)\citenamefont {Kamada}
  \emph {et~al.}}]{Benchmark_2001}%
  \BibitemOpen
  \bibfield  {author} {\bibinfo {author} {\bibfnamefont {H.}~\bibnamefont
  {Kamada}} \emph {et~al.},\ }\href {\doibase 10.1103/PhysRevC.64.044001}
  {\bibfield  {journal} {\bibinfo  {journal} {Phys. Rev.}\ }\textbf {\bibinfo
  {volume} {C64}},\ \bibinfo {pages} {044001} (\bibinfo {year} {2001})},\
  \Eprint {http://arxiv.org/abs/nucl-th/0104057} {arXiv:nucl-th/0104057
  [nucl-th]} \BibitemShut {NoStop}%
%%CITATION = NUCL-TH/0104057;%%
\bibitem [{\citenamefont {Viviani}\ \emph {et~al.}(2017)\citenamefont
  {Viviani}, \citenamefont {Deltuva}, \citenamefont {Lazauskas}, \citenamefont
  {Fonseca}, \citenamefont {Kievsky},\ and\ \citenamefont
  {Marcucci}}]{Benchmark_Hadronic}%
  \BibitemOpen
  \bibfield  {author} {\bibinfo {author} {\bibfnamefont {M.}~\bibnamefont
  {Viviani}}, \bibinfo {author} {\bibfnamefont {A.}~\bibnamefont {Deltuva}},
  \bibinfo {author} {\bibfnamefont {R.}~\bibnamefont {Lazauskas}}, \bibinfo
  {author} {\bibfnamefont {A.~C.}\ \bibnamefont {Fonseca}}, \bibinfo {author}
  {\bibfnamefont {A.}~\bibnamefont {Kievsky}}, \ and\ \bibinfo {author}
  {\bibfnamefont {L.~E.}\ \bibnamefont {Marcucci}},\ }\href {\doibase
  10.1103/PhysRevC.95.034003} {\bibfield  {journal} {\bibinfo  {journal} {Phys.
  Rev. C}\ }\textbf {\bibinfo {volume} {95}},\ \bibinfo {pages} {034003}
  (\bibinfo {year} {2017})}\BibitemShut {NoStop}%
\bibitem [{\citenamefont {Lazauskas}(2018)}]{Lazauskas:2017paz}%
  \BibitemOpen
  \bibfield  {author} {\bibinfo {author} {\bibfnamefont {R.}~\bibnamefont
  {Lazauskas}},\ }\href {\doibase 10.1103/PhysRevC.97.044002} {\bibfield
  {journal} {\bibinfo  {journal} {Phys. Rev.}\ }\textbf {\bibinfo {volume}
  {C97}},\ \bibinfo {pages} {044002} (\bibinfo {year} {2018})},\ \Eprint
  {http://arxiv.org/abs/1711.04716} {arXiv:1711.04716 [nucl-th]} \BibitemShut
  {NoStop}%
%%CITATION = ARXIV:1711.04716;%%
\bibitem [{\citenamefont {{Nevo Dinur}}\ \emph {et~al.}(2014)\citenamefont
  {{Nevo Dinur}}, \citenamefont {Barnea}, \citenamefont {Ji},\ and\
  \citenamefont {Bacca}}]{NND_2014_PRC}%
  \BibitemOpen
  \bibfield  {author} {\bibinfo {author} {\bibfnamefont {N.}~\bibnamefont
  {{Nevo Dinur}}}, \bibinfo {author} {\bibfnamefont {N.}~\bibnamefont
  {Barnea}}, \bibinfo {author} {\bibfnamefont {C.}~\bibnamefont {Ji}}, \ and\
  \bibinfo {author} {\bibfnamefont {S.}~\bibnamefont {Bacca}},\ }\href
  {\doibase 10.1103/PhysRevC.89.064317} {\bibfield  {journal} {\bibinfo
  {journal} {Phys. Rev. C}\ }\textbf {\bibinfo {volume} {89}},\ \bibinfo
  {pages} {064317} (\bibinfo {year} {2014})}\BibitemShut {NoStop}%
\bibitem [{\citenamefont {Baker}\ \emph {et~al.}(2018)\citenamefont {Baker},
  \citenamefont {Launey}, \citenamefont {{Nevo Dinur}}, \citenamefont {Bacca},
  \citenamefont {Draayer},\ and\ \citenamefont {Dytrych}}]{Baker_2018}%
  \BibitemOpen
  \bibfield  {author} {\bibinfo {author} {\bibfnamefont {R.~B.}\ \bibnamefont
  {Baker}}, \bibinfo {author} {\bibfnamefont {K.~D.}\ \bibnamefont {Launey}},
  \bibinfo {author} {\bibfnamefont {N.}~\bibnamefont {{Nevo Dinur}}}, \bibinfo
  {author} {\bibfnamefont {S.}~\bibnamefont {Bacca}}, \bibinfo {author}
  {\bibfnamefont {J.~P.}\ \bibnamefont {Draayer}}, \ and\ \bibinfo {author}
  {\bibfnamefont {T.}~\bibnamefont {Dytrych}},\ }\href {\doibase
  10.1063/1.5078825} {\bibfield  {journal} {\bibinfo  {journal} {AIP Conference
  Proceedings}\ }\textbf {\bibinfo {volume} {2038}},\ \bibinfo {pages} {020006}
  (\bibinfo {year} {2018})},\ \Eprint
  {http://arxiv.org/abs/https://aip.scitation.org/doi/pdf/10.1063/1.5078825}
  {https://aip.scitation.org/doi/pdf/10.1063/1.5078825} \BibitemShut {NoStop}%
\bibitem [{\citenamefont {Hartree}(1958)}]{Numerov}%
  \BibitemOpen
  \bibfield  {author} {\bibinfo {author} {\bibfnamefont {D.~R.}\ \bibnamefont
  {Hartree}},\ }\href@noop {} {\emph {\bibinfo {title} {Numerical Analysis}}}\
  (\bibinfo  {publisher} {Oxford University Press, 2nd ed.},\ \bibinfo {year}
  {1958})\BibitemShut {NoStop}%
\bibitem [{\citenamefont {Hernandez}\ \emph {et~al.}(2018)\citenamefont
  {Hernandez}, \citenamefont {Ekstr\"{o}m}, \citenamefont {{Nevo Dinur}},
  \citenamefont {Ji}, \citenamefont {Bacca},\ and\ \citenamefont
  {Barnea}}]{Hernandez2018}%
  \BibitemOpen
  \bibfield  {author} {\bibinfo {author} {\bibfnamefont {O.}~\bibnamefont
  {Hernandez}}, \bibinfo {author} {\bibfnamefont {A.}~\bibnamefont
  {Ekstr\"{o}m}}, \bibinfo {author} {\bibfnamefont {N.}~\bibnamefont {{Nevo
  Dinur}}}, \bibinfo {author} {\bibfnamefont {C.}~\bibnamefont {Ji}}, \bibinfo
  {author} {\bibfnamefont {S.}~\bibnamefont {Bacca}}, \ and\ \bibinfo {author}
  {\bibfnamefont {N.}~\bibnamefont {Barnea}},\ }\href {\doibase
  10.1016/j.physletb.2018.01.043} {\bibfield  {journal} {\bibinfo  {journal}
  {Physics Letters B}\ }\textbf {\bibinfo {volume} {778}},\ \bibinfo {pages}
  {377 } (\bibinfo {year} {2018})}\BibitemShut {NoStop}%
\bibitem [{\citenamefont {Wiringa}(1991)}]{Wiringa91}%
  \BibitemOpen
  \bibfield  {author} {\bibinfo {author} {\bibfnamefont {R.~B.}\ \bibnamefont
  {Wiringa}},\ }\href {\doibase 10.1103/PhysRevC.43.1585} {\bibfield  {journal}
  {\bibinfo  {journal} {Phys.Rev.}\ }\textbf {\bibinfo {volume} {C43}},\
  \bibinfo {pages} {1585} (\bibinfo {year} {1991})}\BibitemShut {NoStop}%
%%CITATION = PHRVA,C43,1585;%%
\bibitem [{\citenamefont {Pudliner}\ \emph {et~al.}(1997)\citenamefont
  {Pudliner}, \citenamefont {Pandharipande}, \citenamefont {Carlson},
  \citenamefont {Pieper},\ and\ \citenamefont {Wiringa}}]{pudliner1997}%
  \BibitemOpen
  \bibfield  {author} {\bibinfo {author} {\bibfnamefont {B.~S.}\ \bibnamefont
  {Pudliner}}, \bibinfo {author} {\bibfnamefont {V.~R.}\ \bibnamefont
  {Pandharipande}}, \bibinfo {author} {\bibfnamefont {J.}~\bibnamefont
  {Carlson}}, \bibinfo {author} {\bibfnamefont {S.~C.}\ \bibnamefont {Pieper}},
  \ and\ \bibinfo {author} {\bibfnamefont {R.~B.}\ \bibnamefont {Wiringa}},\
  }\href {\doibase 10.1103/PhysRevC.56.1720} {\bibfield  {journal} {\bibinfo
  {journal} {Phys. Rev. C}\ }\textbf {\bibinfo {volume} {56}},\ \bibinfo
  {pages} {1720} (\bibinfo {year} {1997})}\BibitemShut {NoStop}%
\bibitem [{\citenamefont {Piarulli}\ \emph {et~al.}(2013)\citenamefont
  {Piarulli}, \citenamefont {Girlanda}, \citenamefont {Marcucci}, \citenamefont
  {Pastore}, \citenamefont {Schiavilla} \emph {et~al.}}]{Piarulli12}%
  \BibitemOpen
  \bibfield  {author} {\bibinfo {author} {\bibfnamefont {M.}~\bibnamefont
  {Piarulli}}, \bibinfo {author} {\bibfnamefont {L.}~\bibnamefont {Girlanda}},
  \bibinfo {author} {\bibfnamefont {L.~E.}\ \bibnamefont {Marcucci}}, \bibinfo
  {author} {\bibfnamefont {S.}~\bibnamefont {Pastore}}, \bibinfo {author}
  {\bibfnamefont {R.}~\bibnamefont {Schiavilla}},  \emph {et~al.},\ }\href
  {\doibase 10.1103/PhysRevC.87.014006} {\bibfield  {journal} {\bibinfo
  {journal} {Phys.\ Rev.~C}\ }\textbf {\bibinfo {volume} {87}},\ \bibinfo
  {pages} {014006} (\bibinfo {year} {2013})}\BibitemShut {NoStop}%
\bibitem [{\citenamefont {Barnea}\ \emph {et~al.}(2001)\citenamefont {Barnea},
  \citenamefont {Leidemann},\ and\ \citenamefont {Orlandini}}]{Barnea2001}%
  \BibitemOpen
  \bibfield  {author} {\bibinfo {author} {\bibfnamefont {N.}~\bibnamefont
  {Barnea}}, \bibinfo {author} {\bibfnamefont {W.}~\bibnamefont {Leidemann}}, \
  and\ \bibinfo {author} {\bibfnamefont {G.}~\bibnamefont {Orlandini}},\ }\href
  {\doibase http://dx.doi.org/10.1016/S0375-9474(01)00794-1} {\bibfield
  {journal} {\bibinfo  {journal} {Nuclear Physics A}\ }\textbf {\bibinfo
  {volume} {693}},\ \bibinfo {pages} {565 } (\bibinfo {year}
  {2001})}\BibitemShut {NoStop}%
\bibitem [{\citenamefont {Leidemann}\ and\ \citenamefont
  {Orlandini}(2013)}]{Leidemann12}%
  \BibitemOpen
  \bibfield  {author} {\bibinfo {author} {\bibfnamefont {W.}~\bibnamefont
  {Leidemann}}\ and\ \bibinfo {author} {\bibfnamefont {G.}~\bibnamefont
  {Orlandini}},\ }\href {\doibase 10.1016/j.ppnp.2012.09.001} {\bibfield
  {journal} {\bibinfo  {journal} {Prog.Part.Nucl.Phys.}\ }\textbf {\bibinfo
  {volume} {68}},\ \bibinfo {pages} {158} (\bibinfo {year} {2013})}\BibitemShut
  {NoStop}%
%%CITATION = ARXIV:1204.4617;%%
\bibitem [{\citenamefont {Bacca}\ and\ \citenamefont
  {Pastore}(2014)}]{review2014}%
  \BibitemOpen
  \bibfield  {author} {\bibinfo {author} {\bibfnamefont {S.}~\bibnamefont
  {Bacca}}\ and\ \bibinfo {author} {\bibfnamefont {S.}~\bibnamefont
  {Pastore}},\ }\href {\doibase 10.1088/0954-3899/41/12/123002} {\bibfield
  {journal} {\bibinfo  {journal} {J.\ Phys.}\ }\textbf {\bibinfo {volume}
  {G41}},\ \bibinfo {pages} {123002} (\bibinfo {year} {2014})},\ \Eprint
  {http://arxiv.org/abs/1407.3490} {arXiv:1407.3490 [nucl-th]} \BibitemShut
  {NoStop}%
\bibitem [{\citenamefont {Marcucci}\ \emph {et~al.}(2016)\citenamefont
  {Marcucci}, \citenamefont {Gross}, \citenamefont {Pena}, \citenamefont
  {Piarulli}, \citenamefont {Schiavilla}, \citenamefont {Sick}, \citenamefont
  {Stadler}, \citenamefont {Van~Orden},\ and\ \citenamefont
  {Viviani}}]{Marcucci:2015rca}%
  \BibitemOpen
  \bibfield  {author} {\bibinfo {author} {\bibfnamefont {L.~E.}\ \bibnamefont
  {Marcucci}}, \bibinfo {author} {\bibfnamefont {F.}~\bibnamefont {Gross}},
  \bibinfo {author} {\bibfnamefont {M.~T.}\ \bibnamefont {Pena}}, \bibinfo
  {author} {\bibfnamefont {M.}~\bibnamefont {Piarulli}}, \bibinfo {author}
  {\bibfnamefont {R.}~\bibnamefont {Schiavilla}}, \bibinfo {author}
  {\bibfnamefont {I.}~\bibnamefont {Sick}}, \bibinfo {author} {\bibfnamefont
  {A.}~\bibnamefont {Stadler}}, \bibinfo {author} {\bibfnamefont {J.~W.}\
  \bibnamefont {Van~Orden}}, \ and\ \bibinfo {author} {\bibfnamefont
  {M.}~\bibnamefont {Viviani}},\ }\href {\doibase
  10.1088/0954-3899/43/2/023002} {\bibfield  {journal} {\bibinfo  {journal} {J.
  Phys.}\ }\textbf {\bibinfo {volume} {G43}},\ \bibinfo {pages} {023002}
  (\bibinfo {year} {2016})},\ \Eprint {http://arxiv.org/abs/1504.05063}
  {arXiv:1504.05063 [nucl-th]} \BibitemShut {NoStop}%
%%CITATION = ARXIV:1504.05063;%%
\bibitem [{\citenamefont {Carlson}\ and\ \citenamefont
  {Schiavilla}(1998)}]{Carlson:1997qn}%
  \BibitemOpen
  \bibfield  {author} {\bibinfo {author} {\bibfnamefont {J.}~\bibnamefont
  {Carlson}}\ and\ \bibinfo {author} {\bibfnamefont {R.}~\bibnamefont
  {Schiavilla}},\ }\href {\doibase 10.1103/RevModPhys.70.743} {\bibfield
  {journal} {\bibinfo  {journal} {Rev. Mod. Phys.}\ }\textbf {\bibinfo {volume}
  {70}},\ \bibinfo {pages} {743} (\bibinfo {year} {1998})}\BibitemShut
  {NoStop}%
%%CITATION = RMPHA,70,743;%%
\bibitem [{\citenamefont {Kievsky}\ \emph
  {et~al.}(2008{\natexlab{a}})\citenamefont {Kievsky}, \citenamefont {Rosati},
  \citenamefont {Viviani}, \citenamefont {Marcucci},\ and\ \citenamefont
  {Girlanda}}]{Kievsky:2008es}%
  \BibitemOpen
  \bibfield  {author} {\bibinfo {author} {\bibfnamefont {A.}~\bibnamefont
  {Kievsky}}, \bibinfo {author} {\bibfnamefont {S.}~\bibnamefont {Rosati}},
  \bibinfo {author} {\bibfnamefont {M.}~\bibnamefont {Viviani}}, \bibinfo
  {author} {\bibfnamefont {L.~E.}\ \bibnamefont {Marcucci}}, \ and\ \bibinfo
  {author} {\bibfnamefont {L.}~\bibnamefont {Girlanda}},\ }\href {\doibase
  10.1088/0954-3899/35/6/063101} {\bibfield  {journal} {\bibinfo  {journal} {J.
  Phys.}\ }\textbf {\bibinfo {volume} {G35}},\ \bibinfo {pages} {063101}
  (\bibinfo {year} {2008}{\natexlab{a}})},\ \Eprint
  {http://arxiv.org/abs/0805.4688} {arXiv:0805.4688 [nucl-th]} \BibitemShut
  {NoStop}%
%%CITATION = ARXIV:0805.4688;%%
\bibitem [{\citenamefont {Viviani}\ \emph {et~al.}(2006)\citenamefont
  {Viviani}, \citenamefont {Marcucci}, \citenamefont {Rosati}, \citenamefont
  {Kievsky},\ and\ \citenamefont {Girlanda}}]{Viviani:2005gu}%
  \BibitemOpen
  \bibfield  {author} {\bibinfo {author} {\bibfnamefont {M.}~\bibnamefont
  {Viviani}}, \bibinfo {author} {\bibfnamefont {L.~E.}\ \bibnamefont
  {Marcucci}}, \bibinfo {author} {\bibfnamefont {S.}~\bibnamefont {Rosati}},
  \bibinfo {author} {\bibfnamefont {A.}~\bibnamefont {Kievsky}}, \ and\
  \bibinfo {author} {\bibfnamefont {L.}~\bibnamefont {Girlanda}},\ }\href
  {\doibase 10.1007/s00601-006-0158-y} {\bibfield  {journal} {\bibinfo
  {journal} {Few Body Syst.}\ }\textbf {\bibinfo {volume} {39}},\ \bibinfo
  {pages} {159} (\bibinfo {year} {2006})},\ \Eprint
  {http://arxiv.org/abs/nucl-th/0512077} {arXiv:nucl-th/0512077 [nucl-th]}
  \BibitemShut {NoStop}%
%%CITATION = NUCL-TH/0512077;%%
\bibitem [{\citenamefont {Phillips}(2016)}]{DanielAR}%
  \BibitemOpen
  \bibfield  {author} {\bibinfo {author} {\bibfnamefont {D.~R.}\ \bibnamefont
  {Phillips}},\ }\href {\doibase 10.1146/annurev-nucl-102014-022321} {\bibfield
   {journal} {\bibinfo  {journal} {Annual Review of Nuclear and Particle
  Science}\ }\textbf {\bibinfo {volume} {66}},\ \bibinfo {pages} {421}
  (\bibinfo {year} {2016})},\ \Eprint
  {http://arxiv.org/abs/https://doi.org/10.1146/annurev-nucl-102014-022321}
  {https://doi.org/10.1146/annurev-nucl-102014-022321} \BibitemShut {NoStop}%
\bibitem [{\citenamefont {Pastore}\ \emph {et~al.}(2008)\citenamefont
  {Pastore}, \citenamefont {Schiavilla},\ and\ \citenamefont
  {Goity}}]{Pastore08}%
  \BibitemOpen
  \bibfield  {author} {\bibinfo {author} {\bibfnamefont {S.}~\bibnamefont
  {Pastore}}, \bibinfo {author} {\bibfnamefont {R.}~\bibnamefont {Schiavilla}},
  \ and\ \bibinfo {author} {\bibfnamefont {J.~L.}\ \bibnamefont {Goity}},\
  }\href {\doibase 10.1103/PhysRevC.78.064002} {\bibfield  {journal} {\bibinfo
  {journal} {Phys.\ Rev.\ C}\ }\textbf {\bibinfo {volume} {78}},\ \bibinfo
  {pages} {064002} (\bibinfo {year} {2008})}\BibitemShut {NoStop}%
\bibitem [{\citenamefont {Pastore}\ \emph {et~al.}(2009)\citenamefont
  {Pastore}, \citenamefont {Girlanda}, \citenamefont {Schiavilla},
  \citenamefont {Viviani},\ and\ \citenamefont {Wiringa}}]{Pastore09}%
  \BibitemOpen
  \bibfield  {author} {\bibinfo {author} {\bibfnamefont {S.}~\bibnamefont
  {Pastore}}, \bibinfo {author} {\bibfnamefont {L.}~\bibnamefont {Girlanda}},
  \bibinfo {author} {\bibfnamefont {R.}~\bibnamefont {Schiavilla}}, \bibinfo
  {author} {\bibfnamefont {M.}~\bibnamefont {Viviani}}, \ and\ \bibinfo
  {author} {\bibfnamefont {R.~B.}\ \bibnamefont {Wiringa}},\ }\href {\doibase
  10.1103/PhysRevC.80.034004} {\bibfield  {journal} {\bibinfo  {journal}
  {Phys.\ Rev.\ C}\ }\textbf {\bibinfo {volume} {80}},\ \bibinfo {pages}
  {034004} (\bibinfo {year} {2009})}\BibitemShut {NoStop}%
\bibitem [{\citenamefont {Pastore}\ \emph {et~al.}(2011)\citenamefont
  {Pastore}, \citenamefont {Girlanda}, \citenamefont {Schiavilla},\ and\
  \citenamefont {Viviani}}]{Pastore11}%
  \BibitemOpen
  \bibfield  {author} {\bibinfo {author} {\bibfnamefont {S.}~\bibnamefont
  {Pastore}}, \bibinfo {author} {\bibfnamefont {L.}~\bibnamefont {Girlanda}},
  \bibinfo {author} {\bibfnamefont {R.}~\bibnamefont {Schiavilla}}, \ and\
  \bibinfo {author} {\bibfnamefont {M.}~\bibnamefont {Viviani}},\ }\href
  {\doibase 10.1103/PhysRevC.84.024001} {\bibfield  {journal} {\bibinfo
  {journal} {Phys.\ Rev.\ C}\ }\textbf {\bibinfo {volume} {84}},\ \bibinfo
  {pages} {024001} (\bibinfo {year} {2011})}\BibitemShut {NoStop}%
\bibitem [{\citenamefont {Kolling}\ \emph {et~al.}(2009)\citenamefont
  {Kolling}, \citenamefont {Epelbaum}, \citenamefont {Krebs},\ and\
  \citenamefont {Mei\ss{}ner}}]{Kolling09}%
  \BibitemOpen
  \bibfield  {author} {\bibinfo {author} {\bibfnamefont {S.}~\bibnamefont
  {Kolling}}, \bibinfo {author} {\bibfnamefont {E.}~\bibnamefont {Epelbaum}},
  \bibinfo {author} {\bibfnamefont {H.}~\bibnamefont {Krebs}}, \ and\ \bibinfo
  {author} {\bibfnamefont {U.-G.}\ \bibnamefont {Mei\ss{}ner}},\ }\href
  {\doibase 10.1103/PhysRevC.80.045502} {\bibfield  {journal} {\bibinfo
  {journal} {Phys.\ Rev.\ C}\ }\textbf {\bibinfo {volume} {80}},\ \bibinfo
  {pages} {045502} (\bibinfo {year} {2009})}\BibitemShut {NoStop}%
\bibitem [{\citenamefont {Kolling}\ \emph {et~al.}(2011)\citenamefont
  {Kolling}, \citenamefont {Epelbaum}, \citenamefont {Krebs},\ and\
  \citenamefont {Mei\ss{}ner}}]{Kolling11}%
  \BibitemOpen
  \bibfield  {author} {\bibinfo {author} {\bibfnamefont {S.}~\bibnamefont
  {Kolling}}, \bibinfo {author} {\bibfnamefont {E.}~\bibnamefont {Epelbaum}},
  \bibinfo {author} {\bibfnamefont {H.}~\bibnamefont {Krebs}}, \ and\ \bibinfo
  {author} {\bibfnamefont {U.-G.}\ \bibnamefont {Mei\ss{}ner}},\ }\href
  {\doibase 10.1103/PhysRevC.84.054008} {\bibfield  {journal} {\bibinfo
  {journal} {Phys.\ Rev.\ C}\ }\textbf {\bibinfo {volume} {84}},\ \bibinfo
  {pages} {054008} (\bibinfo {year} {2011})}\BibitemShut {NoStop}%
\bibitem [{\citenamefont {Kolling}\ \emph {et~al.}(2012)\citenamefont
  {Kolling}, \citenamefont {Epelbaum},\ and\ \citenamefont
  {Phillips}}]{Kolling12}%
  \BibitemOpen
  \bibfield  {author} {\bibinfo {author} {\bibfnamefont {S.}~\bibnamefont
  {Kolling}}, \bibinfo {author} {\bibfnamefont {E.}~\bibnamefont {Epelbaum}}, \
  and\ \bibinfo {author} {\bibfnamefont {D.~R.}\ \bibnamefont {Phillips}},\
  }\href {\doibase 10.1103/PhysRevC.86.047001} {\bibfield  {journal} {\bibinfo
  {journal} {Phys.\ Rev.}\ }\textbf {\bibinfo {volume} {C86}},\ \bibinfo
  {pages} {047001} (\bibinfo {year} {2012})}\BibitemShut {NoStop}%
%%CITATION = ARXIV:1209.0837;%%
\bibitem [{\citenamefont {Zemach}(1956)}]{Zemach_1956_PR}%
  \BibitemOpen
  \bibfield  {author} {\bibinfo {author} {\bibfnamefont {A.~C.}\ \bibnamefont
  {Zemach}},\ }\href {\doibase 10.1103/PhysRev.104.1771} {\bibfield  {journal}
  {\bibinfo  {journal} {Phys. Rev.}\ }\textbf {\bibinfo {volume} {104}},\
  \bibinfo {pages} {1771} (\bibinfo {year} {1956})}\BibitemShut {NoStop}%
\bibitem [{\citenamefont {Kelly}(2004)}]{Kelly_2004}%
  \BibitemOpen
  \bibfield  {author} {\bibinfo {author} {\bibfnamefont {J.~J.}\ \bibnamefont
  {Kelly}},\ }\href {\doibase 10.1103/PhysRevC.70.068202} {\bibfield  {journal}
  {\bibinfo  {journal} {Phys.\ Rev.\ C}\ }\textbf {\bibinfo {volume} {70}},\
  \bibinfo {pages} {068202} (\bibinfo {year} {2004})}\BibitemShut {NoStop}%
\bibitem [{\citenamefont {Wiringa}\ \emph {et~al.}(1995)\citenamefont
  {Wiringa}, \citenamefont {Stoks},\ and\ \citenamefont {Schiavilla}}]{AV18}%
  \BibitemOpen
  \bibfield  {author} {\bibinfo {author} {\bibfnamefont {R.~B.}\ \bibnamefont
  {Wiringa}}, \bibinfo {author} {\bibfnamefont {V.~G.~J.}\ \bibnamefont
  {Stoks}}, \ and\ \bibinfo {author} {\bibfnamefont {R.}~\bibnamefont
  {Schiavilla}},\ }\href {\doibase 10.1103/PhysRevC.51.38} {\bibfield
  {journal} {\bibinfo  {journal} {Phys.\ Rev.\ C}\ }\textbf {\bibinfo {volume}
  {51}},\ \bibinfo {pages} {38} (\bibinfo {year} {1995})}\BibitemShut {NoStop}%
\bibitem [{\citenamefont {Pudliner}\ \emph {et~al.}(1995)\citenamefont
  {Pudliner}, \citenamefont {Pandharipande}, \citenamefont {Carlson},\ and\
  \citenamefont {Wiringa}}]{Pudliner95}%
  \BibitemOpen
  \bibfield  {author} {\bibinfo {author} {\bibfnamefont {B.~S.}\ \bibnamefont
  {Pudliner}}, \bibinfo {author} {\bibfnamefont {V.~R.}\ \bibnamefont
  {Pandharipande}}, \bibinfo {author} {\bibfnamefont {J.}~\bibnamefont
  {Carlson}}, \ and\ \bibinfo {author} {\bibfnamefont {R.~B.}\ \bibnamefont
  {Wiringa}},\ }\href {\doibase 10.1103/PhysRevLett.74.4396} {\bibfield
  {journal} {\bibinfo  {journal} {Phys.\ Rev.\ Lett.}\ }\textbf {\bibinfo
  {volume} {74}},\ \bibinfo {pages} {4396} (\bibinfo {year}
  {1995})}\BibitemShut {NoStop}%
%%CITATION = PRLTA,74,4396;%%
\bibitem [{\citenamefont {Mohr}\ \emph {et~al.}(2016)\citenamefont {Mohr},
  \citenamefont {Newell},\ and\ \citenamefont {Taylor}}]{CODATA_2014}%
  \BibitemOpen
  \bibfield  {author} {\bibinfo {author} {\bibfnamefont {P.~J.}\ \bibnamefont
  {Mohr}}, \bibinfo {author} {\bibfnamefont {D.~B.}\ \bibnamefont {Newell}}, \
  and\ \bibinfo {author} {\bibfnamefont {B.~N.}\ \bibnamefont {Taylor}},\
  }\href {\doibase 10.1103/RevModPhys.88.035009} {\bibfield  {journal}
  {\bibinfo  {journal} {Rev.\ Mod.\ Phys.}\ }\textbf {\bibinfo {volume} {88}},\
  \bibinfo {pages} {035009} (\bibinfo {year} {2016})}\BibitemShut {NoStop}%
\bibitem [{\citenamefont {Friar}\ and\ \citenamefont
  {Sick}(2004)}]{Friar_2004}%
  \BibitemOpen
  \bibfield  {author} {\bibinfo {author} {\bibfnamefont {J.~L.}\ \bibnamefont
  {Friar}}\ and\ \bibinfo {author} {\bibfnamefont {I.}~\bibnamefont {Sick}},\
  }\href {\doibase http://dx.doi.org/10.1016/j.physletb.2003.11.018} {\bibfield
   {journal} {\bibinfo  {journal} {Physics Letters B}\ }\textbf {\bibinfo
  {volume} {579}},\ \bibinfo {pages} {285} (\bibinfo {year}
  {2004})}\BibitemShut {NoStop}%
\bibitem [{\citenamefont {Afanasev}\ \emph {et~al.}(1998)\citenamefont
  {Afanasev}, \citenamefont {Afanasev},\ and\ \citenamefont
  {Trubnikov}}]{Afanasev_1998}%
  \BibitemOpen
  \bibfield  {author} {\bibinfo {author} {\bibfnamefont {A.~V.}\ \bibnamefont
  {Afanasev}}, \bibinfo {author} {\bibfnamefont {V.~D.}\ \bibnamefont
  {Afanasev}}, \ and\ \bibinfo {author} {\bibfnamefont {S.~V.}\ \bibnamefont
  {Trubnikov}},\ }\href {https://arxiv.org/abs/nucl-th/9808047} {\bibfield
  {journal} {\bibinfo  {journal} {arXiv:nucl-th/9808047}\ } (\bibinfo {year}
  {1998})}\BibitemShut {NoStop}%
\bibitem [{\citenamefont {Sick}(2014)}]{Sick2014}%
  \BibitemOpen
  \bibfield  {author} {\bibinfo {author} {\bibfnamefont {I.}~\bibnamefont
  {Sick}},\ }\href {\doibase 10.1103/PhysRevC.90.064002} {\bibfield  {journal}
  {\bibinfo  {journal} {Phys.\ Rev.\ C}\ }\textbf {\bibinfo {volume} {90}},\
  \bibinfo {pages} {064002} (\bibinfo {year} {2014})}\BibitemShut {NoStop}%
\bibitem [{\citenamefont {Purcell}\ \emph {et~al.}(2010)\citenamefont
  {Purcell}, \citenamefont {Kelley}, \citenamefont {Kwan}, \citenamefont
  {Sheu},\ and\ \citenamefont {Weller}}]{PURCELL20101}%
  \BibitemOpen
  \bibfield  {author} {\bibinfo {author} {\bibfnamefont {J.}~\bibnamefont
  {Purcell}}, \bibinfo {author} {\bibfnamefont {J.}~\bibnamefont {Kelley}},
  \bibinfo {author} {\bibfnamefont {E.}~\bibnamefont {Kwan}}, \bibinfo {author}
  {\bibfnamefont {C.}~\bibnamefont {Sheu}}, \ and\ \bibinfo {author}
  {\bibfnamefont {H.}~\bibnamefont {Weller}},\ }\href {\doibase
  https://doi.org/10.1016/j.nuclphysa.2010.08.012} {\bibfield  {journal}
  {\bibinfo  {journal} {Nuclear Physics A}\ }\textbf {\bibinfo {volume}
  {848}},\ \bibinfo {pages} {1 } (\bibinfo {year} {2010})}\BibitemShut
  {NoStop}%
\bibitem [{\citenamefont {Angeli}\ and\ \citenamefont
  {Marinova}(2013)}]{ANGELI201369}%
  \BibitemOpen
  \bibfield  {author} {\bibinfo {author} {\bibfnamefont {I.}~\bibnamefont
  {Angeli}}\ and\ \bibinfo {author} {\bibfnamefont {K.}~\bibnamefont
  {Marinova}},\ }\href {\doibase https://doi.org/10.1016/j.adt.2011.12.006}
  {\bibfield  {journal} {\bibinfo  {journal} {Atomic Data and Nuclear Data
  Tables}\ }\textbf {\bibinfo {volume} {99}},\ \bibinfo {pages} {69 } (\bibinfo
  {year} {2013})}\BibitemShut {NoStop}%
\bibitem [{\citenamefont {Schiavilla}\ \emph {et~al.}(2018)\citenamefont
  {Schiavilla} \emph {et~al.}}]{Schiavilla:2018udt}%
  \BibitemOpen
  \bibfield  {author} {\bibinfo {author} {\bibfnamefont {R.}~\bibnamefont
  {Schiavilla}} \emph {et~al.},\ }\href@noop {} {\  (\bibinfo {year} {2018})},\
  \Eprint {http://arxiv.org/abs/1809.10180} {arXiv:1809.10180 [nucl-th]}
  \BibitemShut {NoStop}%
%%CITATION = ARXIV:1809.10180;%%
\bibitem [{\citenamefont {Kievsky}\ \emph
  {et~al.}(2008{\natexlab{b}})\citenamefont {Kievsky}, \citenamefont {Rosati},
  \citenamefont {Viviani}, \citenamefont {Marcucci},\ and\ \citenamefont
  {Girlanda}}]{Kievsky_2008_JPG}%
  \BibitemOpen
  \bibfield  {author} {\bibinfo {author} {\bibfnamefont {A.}~\bibnamefont
  {Kievsky}}, \bibinfo {author} {\bibfnamefont {S.}~\bibnamefont {Rosati}},
  \bibinfo {author} {\bibfnamefont {M.}~\bibnamefont {Viviani}}, \bibinfo
  {author} {\bibfnamefont {L.~E.}\ \bibnamefont {Marcucci}}, \ and\ \bibinfo
  {author} {\bibfnamefont {L.}~\bibnamefont {Girlanda}},\ }\href
  {http://stacks.iop.org/0954-3899/35/i=6/a=063101} {\bibfield  {journal}
  {\bibinfo  {journal} {Journal of Physics G: Nuclear and Particle Physics}\
  }\textbf {\bibinfo {volume} {35}},\ \bibinfo {pages} {063101} (\bibinfo
  {year} {2008}{\natexlab{b}})}\BibitemShut {NoStop}%
\bibitem [{\citenamefont {Pieper}\ \emph {et~al.}(2001)\citenamefont {Pieper},
  \citenamefont {Pandharipande}, \citenamefont {Wiringa},\ and\ \citenamefont
  {Carlson}}]{Pieper_2001_PRC}%
  \BibitemOpen
  \bibfield  {author} {\bibinfo {author} {\bibfnamefont {S.~C.}\ \bibnamefont
  {Pieper}}, \bibinfo {author} {\bibfnamefont {V.~R.}\ \bibnamefont
  {Pandharipande}}, \bibinfo {author} {\bibfnamefont {R.~B.}\ \bibnamefont
  {Wiringa}}, \ and\ \bibinfo {author} {\bibfnamefont {J.}~\bibnamefont
  {Carlson}},\ }\href {\doibase 10.1103/PhysRevC.64.014001} {\bibfield
  {journal} {\bibinfo  {journal} {Phys. Rev. C}\ }\textbf {\bibinfo {volume}
  {64}},\ \bibinfo {pages} {014001} (\bibinfo {year} {2001})}\BibitemShut
  {NoStop}%
\bibitem [{\citenamefont {Epelbaum}\ \emph {et~al.}(2015)\citenamefont
  {Epelbaum}, \citenamefont {Krebs},\ and\ \citenamefont
  {Meissner}}]{Epelbaum2015}%
  \BibitemOpen
  \bibfield  {author} {\bibinfo {author} {\bibfnamefont {E.}~\bibnamefont
  {Epelbaum}}, \bibinfo {author} {\bibfnamefont {H.}~\bibnamefont {Krebs}}, \
  and\ \bibinfo {author} {\bibfnamefont {U.-G.}\ \bibnamefont {Meissner}},\
  }\href {http://dx.doi.org/10.1140/epja/i2015-15053-8} {\bibfield  {journal}
  {\bibinfo  {journal} {The European Physical Journal A}\ }\textbf {\bibinfo
  {volume} {51}},\ \bibinfo {pages} {53} (\bibinfo {year} {2015})}\BibitemShut
  {NoStop}%
\bibitem [{\citenamefont {Entem}\ and\ \citenamefont
  {Machleidt}(2003)}]{Entem03}%
  \BibitemOpen
  \bibfield  {author} {\bibinfo {author} {\bibfnamefont {D.~R.}\ \bibnamefont
  {Entem}}\ and\ \bibinfo {author} {\bibfnamefont {R.}~\bibnamefont
  {Machleidt}},\ }\href {\doibase 10.1103/PhysRevC.68.041001} {\bibfield
  {journal} {\bibinfo  {journal} {Phys. Rev. C}\ }\textbf {\bibinfo {volume}
  {68}},\ \bibinfo {pages} {041001} (\bibinfo {year} {2003})}\BibitemShut
  {NoStop}%
\bibitem [{\citenamefont {Navr\'atil}(2007)}]{Navratil07b}%
  \BibitemOpen
  \bibfield  {author} {\bibinfo {author} {\bibfnamefont {P.}~\bibnamefont
  {Navr\'atil}},\ }\href {http://dx.doi.org/10.1007/s00601-007-0193-3}
  {\bibfield  {journal} {\bibinfo  {journal} {Few-Body Systems}\ }\textbf
  {\bibinfo {volume} {41}},\ \bibinfo {pages} {117} (\bibinfo {year}
  {2007})}\BibitemShut {NoStop}%
\bibitem [{\citenamefont {{H\"{o}hler}}\ \emph {et~al.}(1976)\citenamefont
  {{H\"{o}hler}}, \citenamefont {Pietarinen}, \citenamefont {Sabba-Stefanescu},
  \citenamefont {Borkowski}, \citenamefont {Simon}, \citenamefont {Walther},\
  and\ \citenamefont {Wendling}}]{HOHLER1976505}%
  \BibitemOpen
  \bibfield  {author} {\bibinfo {author} {\bibfnamefont {G.}~\bibnamefont
  {{H\"{o}hler}}}, \bibinfo {author} {\bibfnamefont {E.}~\bibnamefont
  {Pietarinen}}, \bibinfo {author} {\bibfnamefont {I.}~\bibnamefont
  {Sabba-Stefanescu}}, \bibinfo {author} {\bibfnamefont {F.}~\bibnamefont
  {Borkowski}}, \bibinfo {author} {\bibfnamefont {G.}~\bibnamefont {Simon}},
  \bibinfo {author} {\bibfnamefont {V.}~\bibnamefont {Walther}}, \ and\
  \bibinfo {author} {\bibfnamefont {R.}~\bibnamefont {Wendling}},\ }\href
  {\doibase https://doi.org/10.1016/0550-3213(76)90449-1} {\bibfield  {journal}
  {\bibinfo  {journal} {Nuclear Physics B}\ }\textbf {\bibinfo {volume}
  {114}},\ \bibinfo {pages} {505 } (\bibinfo {year} {1976})}\BibitemShut
  {NoStop}%
\bibitem [{\citenamefont {Friar}\ and\ \citenamefont
  {Payne}(2005)}]{FriarPayne:2005_PRC-1_NS_HFS}%
  \BibitemOpen
  \bibfield  {author} {\bibinfo {author} {\bibfnamefont {J.~L.}\ \bibnamefont
  {Friar}}\ and\ \bibinfo {author} {\bibfnamefont {G.~L.}\ \bibnamefont
  {Payne}},\ }\href {\doibase 10.1103/PhysRevC.72.014002} {\bibfield  {journal}
  {\bibinfo  {journal} {Phys. Rev. C}\ }\textbf {\bibinfo {volume} {72}},\
  \bibinfo {pages} {014002} (\bibinfo {year} {2005})}\BibitemShut {NoStop}%
\bibitem [{\citenamefont {{J.\ Beringer et al.\ (Particle Data
  Group)}}(2012)}]{PDG_Beringer_2012_PRD}%
  \BibitemOpen
  \bibfield  {author} {\bibinfo {author} {\bibnamefont {{J.\ Beringer et al.\
  (Particle Data Group)}}},\ }\href {\doibase 10.1103/PhysRevD.86.010001}
  {\bibfield  {journal} {\bibinfo  {journal} {Phys. Rev. D}\ }\textbf {\bibinfo
  {volume} {86}},\ \bibinfo {pages} {010001} (\bibinfo {year}
  {2012})}\BibitemShut {NoStop}%
\end{thebibliography}%

\end{document}